# Ab-initio insights into the structural, elastic, bonding, and thermophysical properties of UH$_x$ ($x$ = 1, 2, 3, 5, 6, 7, 8) under pressure: possible relevance to high-$T_c$ superconductivity

*Md. Ashraful Alam$^{1,2}$, F. Parvin$^2$, S. H. Naqib$^{2}$\**

$^1$*Department of Physics, Mawlana Bhashani Science and Technology University, Santosh, Tangail 1902, Bangladesh*

$^2$*Department of Physics, University of Rajshahi, Rajshahi 6205, Bangladesh*

*\*Corresponding* author: Postal address: Department of Physics, University of Rajshahi, Rajshahi 6205, Bangladesh; Cell: +8801715049589; Email*: salehnaqib@yahoo.com*

**Abstract**

Binary uranium hydrides, UH$_x$ ($x$ = 1, 2, 3, 5, 6, 7, 8), with different crystal symmetries are potentially interesting compounds for high-$T_c$ superconductivity and as hydrogen storage systems. In this work we have explored the structural, elastic, mechanical, bonding, and thermophysical properties of these systems under uniform pressure via density functional theory based computations. Most of the results disclosed in this work are novel. From the calculations of the cohesive energy and enthalpy of formation, we have found that the titled compounds are chemically stable. The computed elastic constants at different pressures ensure elastic stability. All the binary hydrides are mechanically anisotropic. Pressure induced brittle-ductile transition takes place under high pressure. The compounds are machinable with the cubic α-UH$_3$-*Pm*-3*n* showing very high value of the machinability index. All the compounds are fairly hard with cubic UH$_8$ showing superhard character. The Debye temperatures and acoustic velocities of these compounds are high; the highest value is found for the cubic UH$_8$. The melting temperature, Grüneisen parameter, minimal phonon thermal conductivity, and the thermal expansion coefficient of these compounds have also been studied at different pressures. All these parameters show excellent correspondence with the estimated Debye temperature, elastic parameters and bonding characteristics.

## 1. Introduction

In 1968 Ashcroft predicted theoretically high temperature superconductivity in solid metallic hydride [1]. From this prediction Sieverts et al. [2] reported UH$_3$ hydride. Later Driggs et al. [3]

and Mulford et al. [4] synthesized and identify $UH_3$ known as the α-$UH_3$ compound. Recently some experimental and theoretical groups predicted several actinium hydrides such as $UH_8$ with high superconducting transition temperature 193 K [5]. Some example of recently reported binary hydrides with superconducting transition temperatures under pressure are: $SiH_4$ [6] ($T_c$ ~17 K @ 96 GPa), $H_3S$ [7, 8] ($T_c$ ~203 K @ 155 GPa), $AcH_5$ [9] ($T_c$ ~79 K @ 150 GPa), $AcH_6$ [9] ($T_c$ ~204 K @ 200 GPa), $AcH_{10}$ [9] ($T_c$ ~251 K @ 200 GPa), $ThH_9$ [9] ($T_c$ ~146 K @ 170 GPa), $ThH_{10}$ [9] ($T_c$ ~161 K @ 170 GPa). Besides experimentally realized and theoretically predicted superconductivity, actinide hydrides have important applications in the nuclear power plant as fuel material [5]. Among the actinide series, uranium is widely used as nuclear material. On the other hand, heated uranium reacts with hydrogen and produces uranium hydrides. The uranium-hydrogen binary system has attracted wide attention of scientific community due to its application in the fields of chemical hydrogen storage material and nuclear technology industry [10]. Uranium hydrides are also candidate of high superconducting temperature materials such as $UH_8$ ($T_c$ ~193 K) [5, 11-15]. For these reasons, recently uranium hydrides have drawn significant attention of the scientific community. At high pressure some new binary uranium hydrides were predicted. Kruglov et al. [16] predicted a series of uranium hydrides such as UH, $U_2H_3$, $UH_2$, $U_2H_5$, $UH_3$, $U_3H_{10}$, $UH_5$, $UH_6$, $U_2H_{13}$, $UH_7$, $UH_8$, $U_2H_{17}$ and $UH_9$ by using ab-initio evolutionary crystal structure prediction. Among of these uranium hydrides, only $UH_7$, $UH_8$ and $UH_9$ have superconducting properties under pressure following theoretical predictions. Wang et al. [17], on the other hand, experimentally confirmed the existence of superconductivity in $UH_5$, $UH_7$ and $UH_8$ under pressure. Souter et al. [18] and Raab et al. [19] investigated and reported results on uranium polyhydrides: UH, $UH_2$, $UH_3$ and $UH_4(H_2)_n$ ($n$ = 0, 1, 2, 3, 4, 5, 6).

Uranium hydrides are novel hydrogen-rich system which contains 5f electrons. Min Liu et al. [5] investigated structural, electronic, mechanical and magnetic properties on some uranium hydride systems. Kruglov et al. [16] predicted crystal structure and calculated $T_c$ for $UH_7$, $UH_8$ and $UH_9$ only for a particular pressure. The calculated values of $T_c$ for $UH_7$ is 65.8 K and 56.7 K (@ 0 GPa), 54.1 K and 43.7 K (@ 20 GPa), for $UH_8$ it is 55.2 K and 46.2 K (@ 0 GPa), 33.3 K and 23.4 K (@ 50 GPa), and for $UH_9$ it is 31.2 K and 19.9 K (@ 300 GPa). Detailed pressure dependent studies of the physical properties of many of these uranium hydrides are still lacking. The superconducting state properties of phonon mediated systems are strongly dependent on the structural and thermophysical properties including the Debye temperature. To address these issues we have selected the UH, UH$_2$, UH$_3$, UH$_5$, UH$_6$, UH$_7$, UH$_8$ binary uranium hydrides in this study. Pressure dependent structural, elastic, bonding and thermophysical properties of these compounds are studied in detail in this work employing density functional theory (DFT) based approach. The pressure ranges have been selected following previous studies [16]. Different structures belonging to the orthorhombic, cubic and hexagonal symmetry have been considered.

The subsequent sections of the work have been arranged as follows: Section 2 describes the methodology briefly. Section 3 contains computational results and discussion. Finally, in Section 4, conclusions are drawn.

**2. Computational scheme**

For geometry optimization of the compounds of interest, plane wave pseudopotential [20] method was used as implemented in the CASTEP code [21]. Generalized gradient approximation (GGA) with PBE functional [22] was used for the electronic exchange-correlation terms. Ultrasoft pseudopotential was used for the calculations of electron-ion interactions [23]. The BFGS algorithm was used to minimize the total energy and internal forces [24] within the optimized crystal structure. Monkhorst–Pack grid [25] was used for $k$-point sampling. The elastic constants were determined using the stress-strain module in CASTEP. Thermophysical parameters were computed from the elastic constants and moduli.

Convergence of tolerance parameters, energy cut-off and size of $k$-point grids of UH$_x$ ($x$ = 1, 2, 3, 5, 6, 7, 8) used for geometry optimizations were as follows:

| Parameters | UH (*Cmcm*) | UH (*P6$_3$/mmc*) | UH$_2$ (*Pbcm*) | α-UH$_3$ (*Pm-3n*) | UH$_5$ (*P6$_3$mc*) | UH$_6$ (*P6$_3$/mmc*) | UH$_7$ (*P6$_3$/mmc*) | UH$_8$ (*Fm-3m*) |
|---|---|---|---|---|---|---|---|---|
| Quality | Ultra-fine | Ultra-fine | Ultra-fine | Ultra-fine | Ultra-fine | Ultra-fine | Ultra-fine | Ultra-fine |
| Energy (eV/atom) | $5.0 \times 10^{-6}$ | $5.0 \times 10^{-6}$ | $5.0 \times 10^{-6}$ | $5.0 \times 10^{-6}$ | $5.0 \times 10^{-6}$ | $5.0 \times 10^{-6}$ | $5.0 \times 10^{-6}$ | $5.0 \times 10^{-6}$ |
| Max. force (eV/Å) | 0.01 | 0.01 | 0.01 | 0.01 | 0.01 | 0.01 | 0.01 | 0.01 |
| Max. stress (GPa) | 0.02 | 0.02 | 0.02 | 0.02 | 0.02 | 0.02 | 0.02 | 0.02 |
| Max. displacement (Å) | $5.0 \times 10^{-4}$ | $5.0 \times 10^{-4}$ | $5.0 \times 10^{-4}$ | $5.0 \times 10^{-4}$ | $5.0 \times 10^{-4}$ | $5.0 \times 10^{-4}$ | $5.0 \times 10^{-4}$ | $5.0 \times 10^{-4}$ |
| Energy cut-off (eV) | 550 | 550 | 550 | 450 | 550 | 550 | 550 | 550 |
| $k$-points | 13×12×17 | 16×16×15 | 7×8×11 | 22×22×22 | 10×10×6 | 15×15×10 | 15×15×9 | 18×18×18 |

## 3. Results and analysis

### 3.1 *Structure and stability*

The different crystal structures of binary uranium hydrides UH$_x$ ($x$ = 1, 2, 3, 5, 6, 7, 8) are shown schematically in Fig. 1. The optimized unit cell parameters (*a*, *b*, *c*, and *V*), the cohesive energy $E_{\text{coh}}$, and the enthalpy of formation ($\Delta H$) at different hydrostatic pressures for each structure are summarized in Table 1.

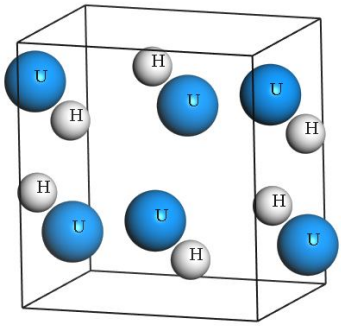

(a) UH-*Cmcm*

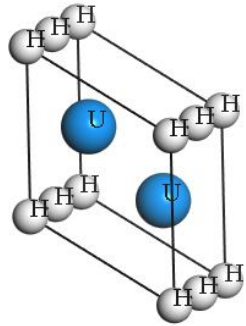

(b) UH-*P6$_3$/mmc*

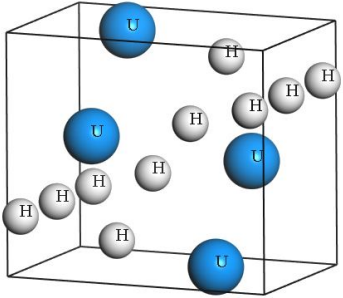

(c) UH$_2$- *Pbcm*

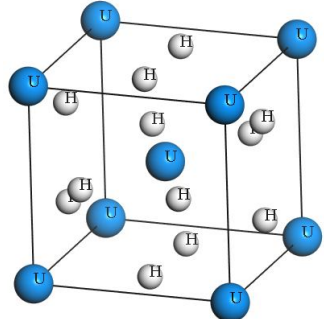

(d) α- UH$_3$-*Pm-3n*

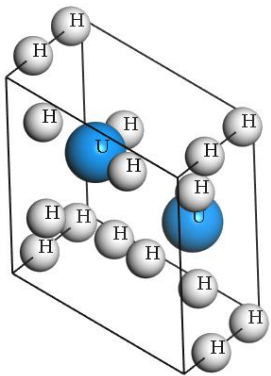

(e) UH$_5$-*P6$_3$mc*

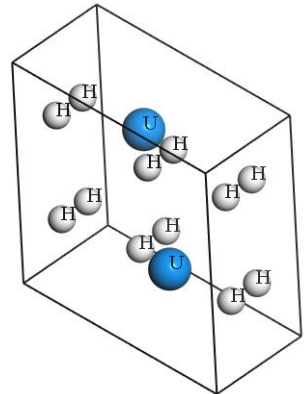

(f) UH$_6$-*P6$_3$/mmc*

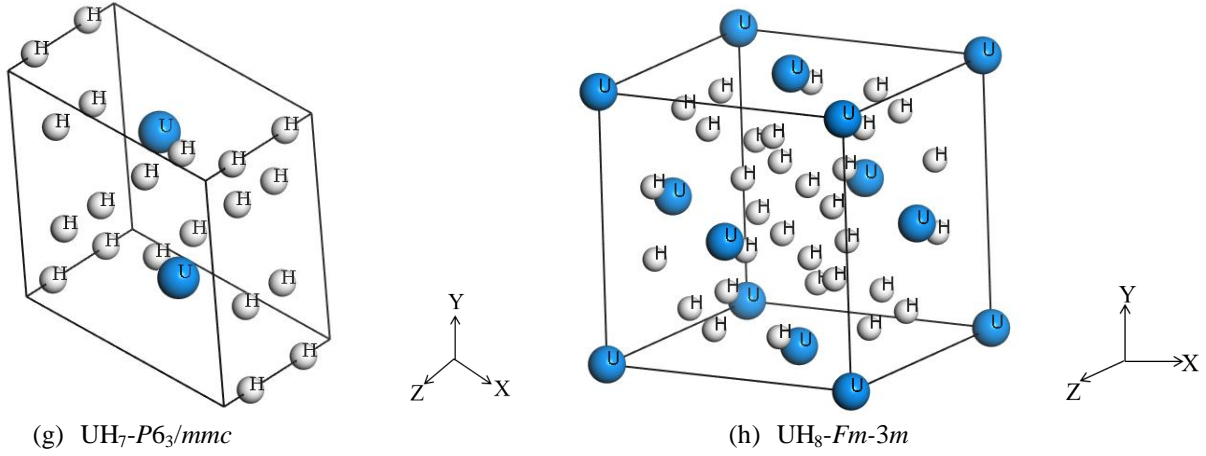

(g) UH$_7$-$P6_3/mmc$   (h) UH$_8$-$Fm$-$3m$

**Figure 1**: Schematic crystal structures of UH$_x$ ($x$ = 1, 2, 3, 5, 6, 7, 8) compounds in different symmetries.

UH-$Cmcm$, UH$_2$-$Pbcm$, and UH$_8$-$Fm$-$3m$ structures contain four formula units and UH-$P6_3/mmc$, α-UH$_3$-$Pm$-$3n$, UH$_5$-$P6_3mc$, UH$_6$-$P6_3/mmc$ and UH$_7$-$P6_3/mmc$ structures contain two formula units in the unit cells. The computed values of the optimized lattice parameters are consistent with the previous results for all of the structures considered herein. In this work, we have calculated the cohesive energy per atom using the approach adopted in Refs. [26-28] for the first time, to determine the chemical stability. The $E_{coh}$ has been computed using the following equation:

$$E_{coh} = \frac{xE_H + E_U - E_{UH_x}}{x+1} \tag{1}$$

where, $E_{UH_x}$ is total energy per formula unit of UH$_x$ and $E_U$ and $E_H$ are the total energies of single U and H atom in the solid states, respectively, and $x$ is the number of H atoms in formula. From Table 1, it is found that the values of cohesive energy per atom are positive, which indicate that all of the structures of UH$_x$ ($x$ = 1, 2, 3, 5, 6, 7, 8) are chemically stable.

**Table 1**. Structural properties of UH$_x$ ($x$ = 1, 2, 3, 5, 6, 7, 8) compounds under different pressures.

| Compounds | Crystal System Space Group Space Group No. | Pressure, $P$ (GPa) | Lattice parameters (Å) | | | Deviation (%) from earlier study [16] | | | Volume, $V$ (Å$^3$) | Cohesive energy (eV/atom) | Enthalpy, ($\Delta H$) (eV) ×10$^3$ |
|---|---|---|---|---|---|---|---|---|---|---|---|
| | | | $a$ | $b$ | $c$ | $a$ | $b$ | $c$ | | | |
| UH | Orthorhombic | 20 | 4.73550 | 5.00747 | 3.58450 | | | | 85.00 | 6.27 | -5.66860 |
| | *Cmcm* | 30 | 4.64283 | 4.95360 | 3.54750 | 2.08 | 1.25 | 1.02 | 81.59 | 6.21 | -5.66340 |
| | 63 | 30 | 4.63700 | 4.94500 | 3.54800 | | | | | -- | |
| | | 40 | 4.56404 | 4.90867 | 3.51432 | | | | 78.73 | 6.13 | -5.65840 |
| UH | Hexagonal | 55 | 3.50607 | 3.50607 | 3.39310 | | | | 36.12 | 5.86 | -2.82556 |
| | *P6$_3$/mmc* | 65 | 3.47707 | 3.47707 | 3.36195 | -0.03 | -0.03 | 0.03 | 35.20 | 5.78 | -2.82333 |
| | 194 | 65 | 3.47800 | 3.47800 | 3.36100 | | | | | | |
| | | 75 | 3.45106 | 3.45106 | 3.33508 | | | | 34.40 | 5.69 | -2.82116 |
| UH$_2$ | Orthorhombic | 20 | 5.53594 | 4.90788 | 3.63823 | | | | 98.85 | 5.40 | -5.73136 |
| | *Pbcm* | 30 | 5.42330 | 4.86632 | 3.59109 | 0.47 | -0.01 | 0.06 | 94.77 | 5.35 | -5.72532 |
| | 57 | 30 | 5.39800 | 4.86700 | 3.58900 | | | | | | |
| | | 40 | 5.32447 | 4.83174 | 3.54986 | | | | 91.33 | 5.29 | -5.71952 |
| α-UH$_3$ | Cubic | 0 | 4.04861 | 4.04861 | 4.04861 | -1.71 | -1.71 | - | 66.36 | 5.02 | -2.90452 |
| | *Pm-3n* | 0 | 4.11800 | 4.11800 | 4.11800 | | | | | | -- |
| | 223 | 10 | 3.95685 | 3.95685 | 3.95685 | | | | 61.95 | 5.01 | -2.90052 |
| | | 20 | 3.88494 | 3.88494 | 3.88494 | | | | 58.63 | 4.97 | -2.89676 |
| UH$_5$ | Hexagonal | 10 | 3.75315 | 3.75315 | 5.90715 | | | | 72.06 | 4.40 | -2.96261 |
| | *P6$_3$mc* | 20 | 3.69378 | 3.69378 | 5.79482 | - | - | - | 68.47 | 4.37 | -2.95823 |
| | 186 | 20 | 3.69500 | 3.69500 | 5.81600 | | | | | | -- |
| | | 30 | 3.64546 | 3.64546 | 5.69702 | | | | 65.57 | 4.33 | -2.95405 |
| UH$_6$ | Hexagonal | 10 | 4.02046 | 4.02046 | 5.44992 | | | | 76.29 | 4.21 | -2.99338 |
| | *P6$_3$/mmc* | 20 | 3.94806 | 3.94806 | 5.37774 | 0.10 | 0.10 | - | 72.59 | 4.19 | -2.98874 |
| | 194 | 20 | 3.94400 | 3.94400 | 5.38500 | | | | | | -- |
| | | 30 | 3.89010 | 3.89010 | 5.31418 | | | | 69.64 | 4.15 | -2.98431 |
| | Hexagonal | 0 | 4.04069 | 4.04069 | 5.83566 | | | | 82.51 | 4.04 | -3.02874 |
| UH$_7$ | *P6$_3$/mmc* | 10 | 3.96083 | 3.96083 | 5.73643 | | | | 77.94 | 4.03 | -3.02374 |
| | 194 | 20 | 3.89809 | 3.89809 | 5.65426 | 0.10 | 0.10 | 0.16 | 74.41 | 4.01 | -3.01899 |
| | | 20 | 3.89400 | 3.89400 | 5.64500 | | | | | | -- |
| | | 30 | 3.84427 | 3.84427 | 5.58619 | | | | 71.49 | 3.98 | -3.01444 |
| | | 40 | 3.79888 | 3.79888 | 5.52278 | | | | 69.02 | 3.95 | -3.01006 |
| UH$_8$ | Cubic | 0 | 5.53373 | 5.53373 | 5.53373 | | | | 169.45 | 3.85 | -6.11670 |
| | *Fm-3m* | 30 | 5.27966 | 5.27966 | 5.27966 | | | | 147.17 | 3.80 | -6.08729 |
| | 225 | 40 | 5.22016 | 5.22016 | 5.22016 | | | | 142.25 | 3.77 | -6.08720 |
| | | 50 | 5.16651 | 5.16651 | 5.16651 | 0.16 | 0.16 | 0.16 | 137.91 | 3.73 | -6.06952 |
| | | 50 | 5.15800 | 5.15800 | 5.15800 | | | | | | |
| | | 60 | 5.11837 | 5.11837 | 5.11837 | | | | 134.09 | 3.70 | -6.06103 |

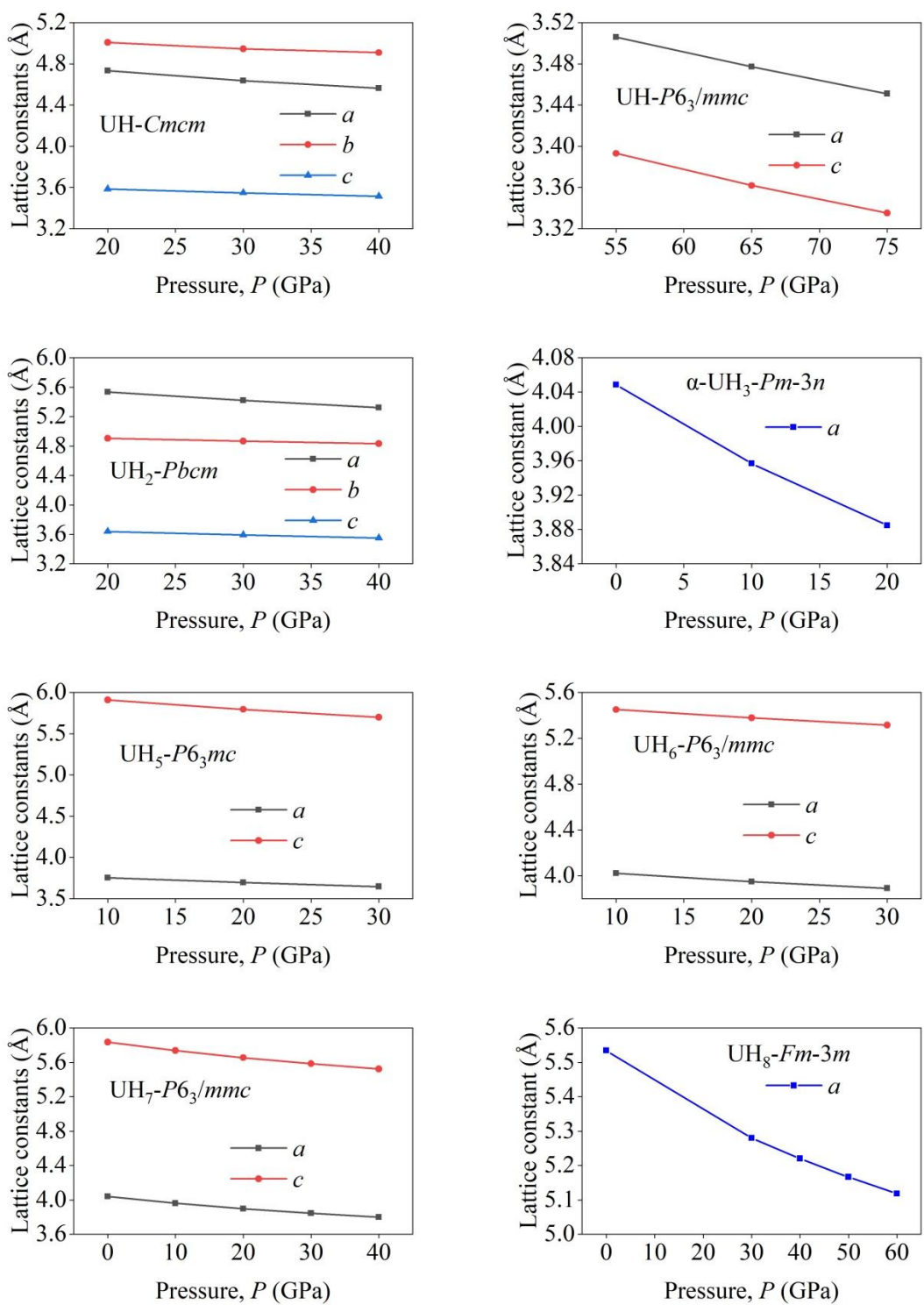

**Figure 2**: Lattice constants of $UH_x$ ($x$ = 1, 2, 3, 5, 6, 7, 8) compounds under pressure.

For all the compounds under consideration, the values of lattice parameters *a*, *b* and *c* decrease with pressure shown in Table 1 (cf. Fig. 2). We do not see any abrupt change in the lattice parameters with, which indicates that the structures are stable in the pressure range considered. This is consistent with the predictions of earlier [16, 17]. Among the binaries, the cubic UH$_8$-*Fm-3m* structure shows the highest axial compressibility. This particular compound also has significantly larger unit cell volume.

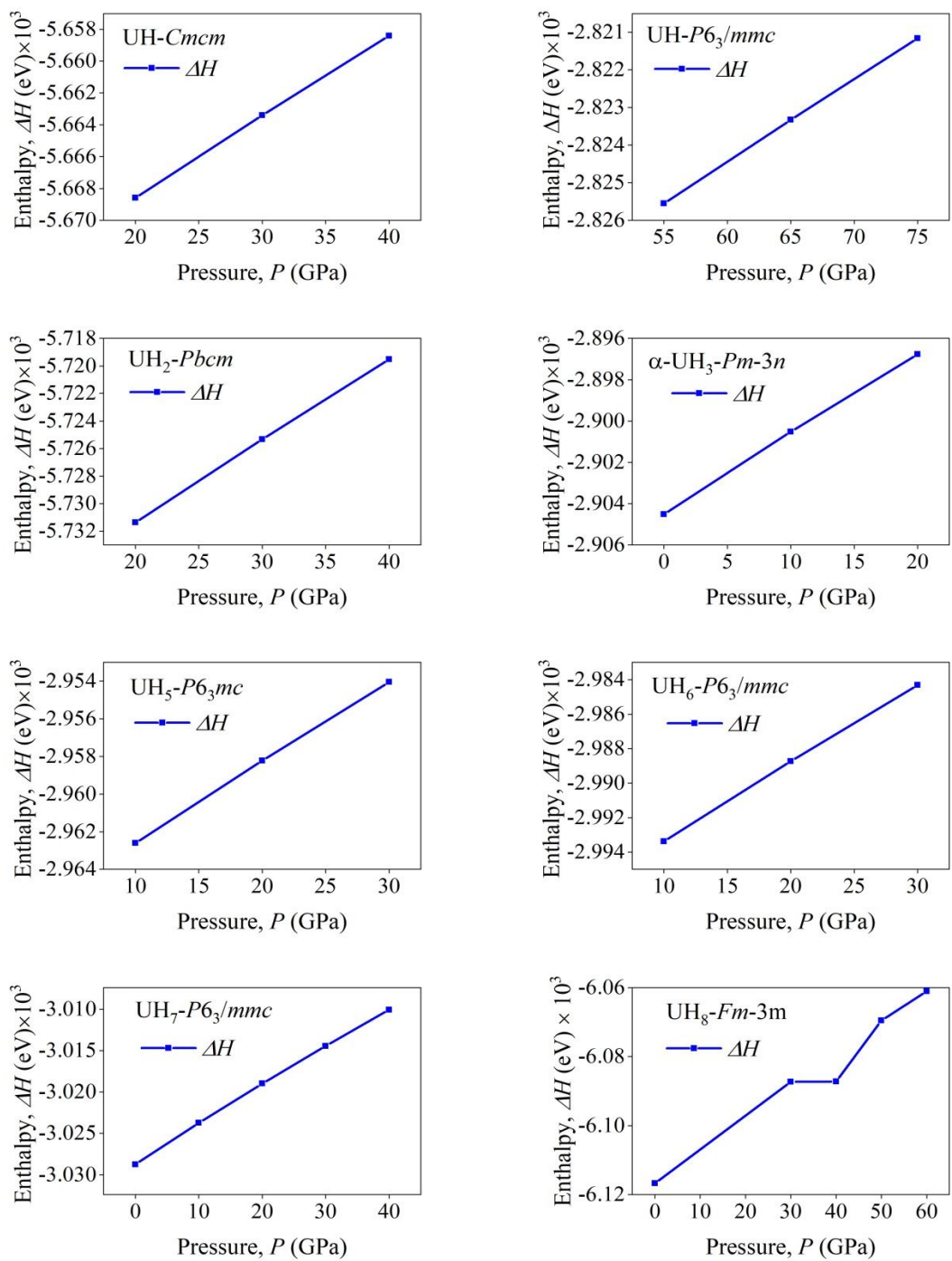

**Figure 3**: Enthalpy of formation of $UH_x$ ($x$ = 1, 2, 3, 5, 6, 7, 8) compounds under pressure.

The values of enthalpies of formation under pressure support the phase stability (cf. Fig. 3) of the $UH_x$ ($x$ = 1, 2, 3, 5, 6, 7, 8) compounds under different hydrostatic pressures. The increase in the enthalpy with increase in pressure is natural since both the internal energy and the product $PV$ increase with pressure for solids. The nonmonotonic behavior observed for cubic $UH_8$-*Fm-3m* in the pressure range 30 GPa – 40 GPa implies that the conventional variations of the internal energy and the product *PV* are not maintained in this pressure interval. This may imply a change in the bonding features at these pressures.

*3.2 Elastic properties*

*3.2.1 Single crystal elastic constants*

For the orthorhombic crystal structure there are nine independent elastic constants: $C_{11}$, $C_{22}$, $C_{33}$, $C_{44}$, $C_{55}$, $C_{66}$, $C_{12}$, $C_{13}$ and $C_{23}$. For hexagonal systems there are six elastic constants: $C_{11}$, $C_{12}$, $C_{13}$, $C_{33}$, $C_{44}$, and $C_{66}$. Except $C_{66}$, the other five are independent elastic constants in this case. For cubic crystal there are three independent elastic constants: $C_{11}$, $C_{12}$, and $C_{44}$. All of these elastic constants are tabulated in Table 2 for the $UH_x$ (*x* = 1, 2, 3, 5, 6, 7, 8) compounds at different pressures. From Table 2, it is seen that $C_{ij}$ increases almost linearly with pressure. All the computed values of $C_{ij}$ are positive and all these parameters show monotonic variation with pressure indicating that there are no pressure induced structural modifications in the binary uranium hydrides under study for the pressure ranges considered. The values of $C_{11}$, $C_{22}$, and $C_{33}$ represent along [100], [010] and [001] directional resistance against uniaxial stress, respectively. Greater values of $C_{11}$, $C_{22}$, and $C_{33}$ indicate higher levels of incompressibility in the corresponding directions. From Table 2, it is observed that UH-*Cmcm* has the highest incompressibility along [001] direction because $C_{33} > C_{22} > C_{11}$. UH-*P6_3/mmc* has the highest incompressibility along [001] directions because $C_{33} > C_{22} = C_{11}$. $UH_2$-*Pbcm* has highest incompressibility along [010] directions because $C_{22} > C_{33} > C_{11}$. For α-$UH_3$-*Pm-3n* incompressibility is the same along the three principle axes because cubic symmetry demands $C_{11} = C_{22} = C_{33}$. The most incompressible direction is [001] for $UH_5$-*P6_3mc*, $UH_6$-*P6_3/mmc*, and for $UH_7$-*P6_3/mmc*, it is [001] because $C_{33} > C_{22} = C_{11}$. The largest values of the elastic constants are found in the UH-*P6_3/mmc* and $UH_8$-*Fm-3m* structures indicating that directional deformation resistances are very strong in these two compounds. The elastic stability of the compounds can be judged from the values of $C_{ij}$.

**Table 2**. Calculated single crystal elastic constants $C_{ij}$ (GPa) of UH$_x$ ($x$ = 1, 2, 3, 5, 6, 7, 8) compounds under different pressures.

| Compounds | Crystal System Space Group Space Group No. | Pressure, $P$ (GPa) | Single crystal elastic constants, $C_{ij}$ (GPa) | | | | | | | | | Ref. |
|---|---|---|---|---|---|---|---|---|---|---|---|---|
| | | | $C_{11}$ | $C_{12}$ | $C_{13}$ | $C_{22}$ | $C_{23}$ | $C_{33}$ | $C_{44}$ | $C_{55}$ | $C_{66}$ | |
| UH | Orthorhombic | 20 | 357.75 | 110.19 | 87.80 | 460.21 | 188.42 | 527.39 | 201.28 | 87.15 | 87.67 | |
| | *Cmcm* | 30 | 413.51 | 142.09 | 101.22 | 524.19 | 222.31 | 600.52 | 222.68 | 94.81 | 92.19 | [This] |
| | 63 | 40 | 473.87 | 173.63 | 112.37 | 590.52 | 257.46 | 669.77 | 244.99 | 102.44 | 86.19 | |
| UH | Hexagonal | 55 | 743.91 | 276.35 | 112.24 | 743.91 | 112.24 | 812.60 | 123.25 | 123.25 | 233.78 | |
| | $P6_3/mmc$ | 65 | 812.88 | 300.54 | 142.71 | 812.88 | 142.71 | 904.98 | 134.88 | 134.88 | 256.17 | |
| | 194 | 75 | 880.27 | 330.09 | 172.47 | 880.27 | 172.47 | 987.28 | 147.83 | 147.83 | 275.09 | |
| UH$_2$ | Orthorhombic | 20 | 348.35 | 136.33 | 90.70 | 506.76 | 165.57 | 456.25 | 170.51 | 76.45 | 73.35 | |
| | *Pbcm* | 30 | 385.88 | 158.77 | 107.21 | 575.02 | 201.71 | 522.22 | 199.70 | 71.70 | 84.34 | [This] |
| | 57 | 40 | 417.07 | 181.11 | 122.23 | 635.98 | 238.85 | 584.59 | 224.88 | 68.01 | 96.01 | |
| α-UH$_3$ | Cubic | 0 | 286.26 | 46.52 | 46.52 | 286.26 | 46.52 | 286.26 | 63.64 | 63.64 | 63.64 | |
| | *Pm-3n* | 10 | 351.56 | 71.07 | 71.07 | 351.56 | 71.07 | 351.56 | 55.43 | 55.43 | 55.43 | [This] |
| | 223 | 20 | 416.39 | 90.17 | 90.17 | 416.39 | 90.17 | 416.39 | 34.16 | 34.16 | 34.16 | |
| UH$_5$ | Hexagonal | 10 | 308.60 | 147.78 | 86.45 | 308.60 | 86.45 | 318.64 | 104.95 | 104.95 | 80.41 | |
| | $P6_3mc$ | 20 | 369.73 | 177.98 | 115.12 | 369.73 | 115.12 | 364.82 | 109.93 | 109.93 | 95.88 | [This] |
| | 186 | 30 | 426.74 | 208.46 | 139.50 | 426.74 | 139.50 | 406.16 | 111.71 | 111.71 | 109.14 | |
| UH$_6$ | Hexagonal | 10 | 331.05 | 101.31 | 92.27 | 331.05 | 92.27 | 440.92 | 122.61 | 122.61 | 114.87 | |
| | $P6_3/mmc$ | 20 | 387.06 | 131.48 | 119.13 | 387.06 | 119.13 | 498.68 | 134.33 | 134.33 | 127.79 | [This] |
| | 194 | 30 | 441.84 | 161.02 | 143.29 | 441.84 | 143.29 | 551.63 | 145.77 | 145.77 | 140.41 | |
| UH$_7$ | Hexagonal | 0 | 310.90 | 74.70 | 65.07 | 310.90 | 65.07 | 371.61 | 132.27 | 132.27 | 118.10 | |
| | $P6_3/mmc$ | 10 | 375.24 | 100.36 | 91.26 | 375.24 | 91.26 | 431.00 | 151.96 | 151.96 | 137.44 | |
| | 194 | 20 | 438.75 | 125.07 | 118.36 | 438.75 | 118.36 | 489.23 | 167.80 | 167.80 | 156.84 | [This] |
| | | 30 | 494.34 | 149.85 | 146.00 | 494.34 | 146.00 | 526.78 | 182.63 | 182.63 | 172.24 | |
| | | 40 | 548.15 | 174.44 | 171.85 | 548.15 | 171.85 | 588.04 | 197.26 | 197.26 | 186.85 | |
| UH$_8$ | Cubic | 0 | 373.27 | 53.54 | 53.54 | 373.27 | 53.54 | 373.27 | 133.70 | 133.70 | 133.70 | |
| | *Fm-3m* | 30 | 603.30 | 104.23 | 104.23 | 603.30 | 104.23 | 603.30 | 184.16 | 184.16 | 184.16 | |
| | 225 | 40 | 682.30 | 127.99 | 127.99 | 682.30 | 127.99 | 682.30 | 198.92 | 198.92 | 198.92 | |
| | | 50 | 731.74 | 143.45 | 143.45 | 731.74 | 143.45 | 731.74 | 213.15 | 213.15 | 213.15 | [This] |
| | | 60 | 795.15 | 161.83 | 161.83 | 795.15 | 161.83 | 795.15 | 226.99 | 226.99 | 226.99 | |

Elastic/mechanical stability conditions of crystals with different structures under pressure are as follows [29-31].

For the orthorhombic structure:

$$C_{11} - P > 0, C_{22} - P > 0, C_{33} - P > 0, C_{44} - P > 0, C_{55} - P > 0, C_{66} - P > 0 \qquad (2)$$

$$and\ (C_{11} + C_{22} - 2C_{12} - 4P) > 0, (C_{11} + C_{33} - 2C_{13} - 4P) > 0,$$

$$(C_{22} + C_{33} - 2C_{23} - 4P) > 0,$$

$$(C_{11} + C_{22} + C_{33} + 2C_{12} + 2C_{13} + 2C_{23} + 3P) > 0 \qquad (3)$$

For the hexagonal structure:

$$\tilde{C}_{11} > 0;\ \tilde{C}_{33} > 0;\ \tilde{C}_{44} > 0;\ (\tilde{C}_{11} - \tilde{C}_{12}) > 0;\ (\tilde{C}_{11} + \tilde{C}_{12})\tilde{C}_{33} > 2\tilde{C}_{13}^2 \qquad (4)$$

where, $\tilde{C}_{\alpha\alpha} = C_{\alpha\alpha} - P, \alpha\alpha = 11, 33, 44, \tilde{C}_{12} = C_{12} + P, \tilde{C}_{13} = C_{13} + P$

For the cubic structure:

$$M_1 = \frac{C_{11} + 2C_{12} + P}{3} > 0;\ M_2 = C_{44} - P > 0;\ M_3 = \frac{C_{11} - 2C_{12} - 2P}{2} > 0 \qquad (5)$$

Our calculated results show that all the UH$_x$ ($x$ = 1, 2, 3, 5, 6, 7, 8) compounds with different structures satisfy the elastic stability conditions under pressure. Besides the calculated values of the tetragonal shear moduli are also positive (Table 3) which indicates that the structures are expected to be dynamically stable as well [32]. Results disclosed in Tables 2 and 3 are novel and no comparison with published sources can be made.

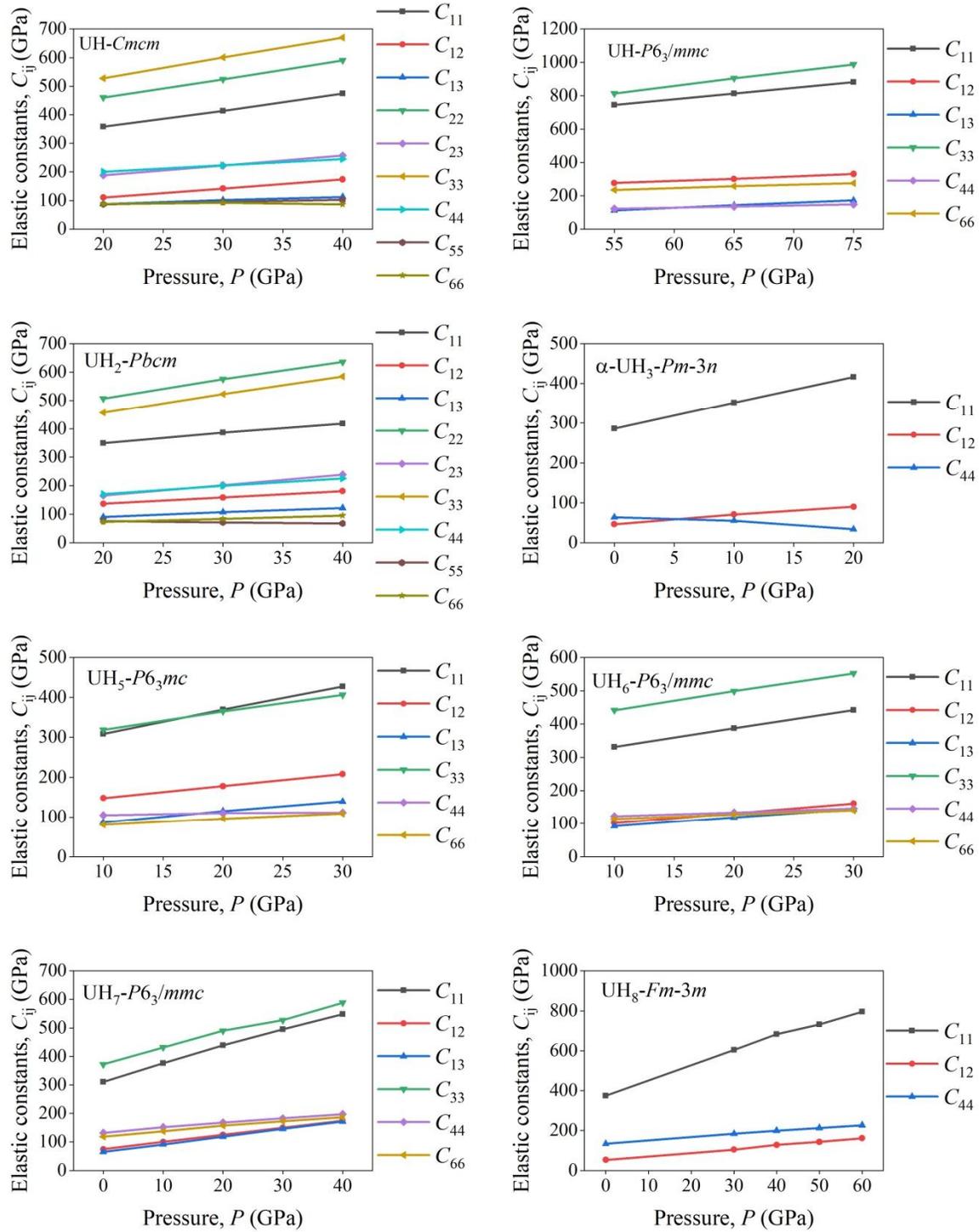

**Figure 4**: Elastic constants $C_{ij}$ of $UH_x$ ($x$ = 1, 2, 3, 5, 6, 7, 8) compounds under different pressures.

**Table 3**. Calculated Cauchy pressure (GPa) and tetragonal shear modulus (TSM) (GPa) of $UH_x$ ($x$ = 1, 2, 3, 5, 6, 7, 8) compounds under different pressures.

| Compounds | Crystal System Space Group Space Group No. | Pressure, $P$ (GPa) | Cauchy Pressure (GPa) | | | TSM (GPa) | Ref. |
|---|---|---|---|---|---|---|---|
| | | | $P_C^a$ | $P_C^b$ | $P_C^c$ | | |
| UH | Orthorhombic | 20 | -12.86 | 0.65 | 22.51 | 123.78 | |
| | *Cmcm* | 30 | -0.37 | 6.42 | 49.90 | 135.71 | [This] |
| | 63 | 40 | 12.47 | 9.94 | 87.44 | 150.12 | |
| UH | Hexagonal | 55 | -11.01 | 42.57 | | 233.78 | |
| | P6$_3$/mmc | 65 | 7.83 | 44.37 | | 256.17 | [This] |
| | 194 | 75 | 24.64 | 55.00 | | 275.09 | |
| UH$_2$ | Orthorhombic | 20 | -4.94 | 14.25 | 62.98 | 106.01 | |
| | *Pbcm* | 30 | 2.01 | 35.52 | 74.43 | 113.56 | [This] |
| | 57 | 40 | 13.97 | 54.21 | 85.09 | 117.98 | |
| α-UH$_3$ | Cubic | 0 | | | -17.11 | 119.87 | |
| | *Pm-3n* | 10 | | | 15.64 | 140.24 | [This] |
| | 223 | 20 | | | 56.01 | 163.11 | |
| UH$_5$ | Hexagonal | 10 | -18.50 | 67.37 | | 80.41 | |
| | *P6$_3$mc* | 20 | 5.19 | 82.10 | | 95.88 | [This] |
| | 186 | 30 | 27.79 | 99.32 | | 109.14 | |
| UH$_6$ | Hexagonal | 10 | -30.35 | -13.57 | | 114.87 | |
| | *P6$_3$/mmc* | 20 | -15.20 | 3.69 | | 127.79 | [This] |
| | 194 | 30 | -2.48 | 20.61 | | 140.41 | |
| UH$_7$ | Hexagonal | 0 | -67.20 | -43.40 | | 118.10 | |
| | *P6$_3$/mmc* | 10 | -60.70 | -37.08 | | 137.44 | |
| | 194 | 20 | -49.44 | -31.77 | | 156.84 | [This] |
| | | 30 | -36.62 | -22.40 | | 172.24 | |
| | | 40 | -25.41 | -12.41 | | 186.85 | |
| UH$_8$ | Cubic | 0 | | | -80.16 | 159.87 | |
| | *Fm-3m* | 30 | | | -79.92 | 249.53 | |
| | 225 | 40 | | | -70.92 | 277.15 | |
| | | 50 | | | -69.71 | 294.15 | [This] |
| | | 60 | | | -65.16 | 316.66 | |

The calculated values of Cauchy pressures ($P_C$) and tetragonal shear modulus are given in Table 3. Cauchy pressure ($P_C$) is used to describe the brittleness and ductility of the compounds. Cauchy pressure ($P_C$) is defined for different crystal symmetries [32] as follows:

For orthorhombic symmetry: $P_C^a = (C_{23} - C_{44})$, $P_C^b = (C_{13} - C_{55})$ and $P_C^c = (C_{12} - C_{66})$

For hexagonal symmetry: $P_C^a = (C_{13} - C_{44})$ and $P_C^b = (C_{12} - C_{66})$

For cubic symmetry: $P_C^c = (C_{12} - C_{44})$

The values of $P_C$ are negative for covalent materials with brittle nature (i.e. $P_C < 0$). Positive Cauchy pressure, on the other hand, suggests ductility with dominant ionic/metallic bonding where the outer shell electrons are almost delocalized [33]. The signs of at least one of the $P_C$ are negative for UH-$Cmcm$ at 20 and 30 GPa, for UH-$P6_3/mmc$ at 55 GPa, for UH$_2$-$Pbcm$ at 20 GPa, for α-UH$_3$-$Pm$-$3n$ at 0 GPa, for UH$_5$-$P6_3mc$ at 10 GPa, for UH$_6$-$P6_3/mmc$ in the pressure range 10-30 GPa, for UH$_7$-$P6_3/mmc$ in the pressure range 0-40 GPa, and for UH$_8$ in the pressure range 0-60 GPa. In some of the UH$_x$ ($x$ = 1, 2, 3, 5, 6, 7, 8) compounds, the negative Cauchy pressures take up positive values at elevated temperature and therefore, imply brittle-ductile transition. This could be related to increased level of metallicity in the UH$_x$ materials with increasing pressure. The value of TSM used to measure the shear stiffness of a crystal i.e. the resistance to shear stress or shear deformation. TSM is also linked with the sound velocity in the solid. Positive values of TSM are indicative of dynamical stability of the structure. The polycrystalline elastic properties, namely the bulk modulus ($B$), shear modulus ($G$) and Young's modulus ($Y$) are given in Table 4. The values of Voigt [34], Reuss [35] and Hill [36] approximations are denoted by the subscripts $V$, $R$ and $H$, respectively. Further details regarding the VRH approach can be found elsewhere [33-36]. Hill approximated elastic moduli are the average of the Voigt and Reuss approximated elastic moduli. From Table 4 (cf. Fig. 5) it is observed that elastic moduli increase systematically with pressure of all the UH$_x$ compounds except the cubic α-UH$_3$-$Pm$-$3n$.

The bulk modulus represents the elastic resistance against uniform volume deformation. Large value of $B$ signifies high incompressibility. Bulk modulus is also an average measure of the strength of interatomic bondings. From Table 4 it is observed that the bulk modulus increases with pressure for all the uranium hydrides under study. The shear modulus represents the elastic resistance against shape changing stress. High shear modulus is indicative of directional chemical bonding. From Table 4 it is found that the resistance to shape change increases with pressure of all the binary hydrides, except the cubic α-UH$_3$-$Pm$-$3n$ compound. This exception is also found in case of the Young's modulus which represents the resistance against tensile stress. Young's modulus is a useful indicator of the stiffness of a solid. Thus it appears that, contrary to conventional behavior where elastic moduli increase with increasing pressure, the situation is reversed except in the cubic α-UH$_3$-$Pm$-$3n$. A possible reason might be that as pressure is increased the bonding characters are modified and become less angular in nature. It is interesting to observe that, for UH, the hexagonal structure has significantly higher elastic moduli compared to the cubic one. This could be a consequence of greater packing fraction in the hexagonal structure.

**Table 4**. Polycrystalline elastic properties of $UH_x$ ($x$ = 1, 2, 3, 5, 6, 7, 8) compounds under different pressures.

| Compounds | Crystal System Space Group Space Group No. | Pressure, $P$ (GPa) | Polycrystalline elastic moduli (GPa) ||||||||| Ref. |
| | | | Bulk, $B$ (GPa) ||| Shear, $G$ (GPa) ||| Young, $Y$ (GPa) ||| |
| | | | $B_V$ | $B_R$ | $B_H$ | $G_V$ | $G_R$ | $G_H$ | $Y_V$ | $Y_R$ | $Y_H$ | |
| UH | Orthorhombic | 20 | 235.35 | 224.00 | 229.68 | 139.15 | 122.50 | 130.83 | 348.72 | 310.84 | 329.85 | |
| | *Cmcm* | 30 | 274.38 | 262.09 | 268.24 | 153.44 | 133.20 | 143.32 | 388.00 | 341.72 | 364.97 | [This] |
| | 63 | 40 | 313.45 | 300.46 | 306.96 | 166.10 | 138.90 | 152.50 | 423.50 | 361.07 | 392.51 | |
| UH | Hexagonal | 55 | 366.90 | 365.97 | 366.44 | 216.03 | 181.68 | 198.85 | 541.76 | 467.65 | 505.18 | |
| | $P6_3/mmc$ | 65 | 411.41 | 411.00 | 411.20 | 234.84 | 198.51 | 216.67 | 591.90 | 512.94 | 552.91 | [This] |
| | 194 | 75 | 455.32 | 455.10 | 455.21 | 252.34 | 215.47 | 233.90 | 638.97 | 558.29 | 599.09 | |
| $UH_2$ | Orthorhombic | 20 | 232.95 | 222.46 | 227.70 | 125.31 | 108.71 | 117.01 | 318.78 | 280.44 | 299.69 | |
| | *Pbcm* | 30 | 268.72 | 254.53 | 261.63 | 138.84 | 116.26 | 127.55 | 355.33 | 302.71 | 329.17 | [This] |
| | 57 | 40 | 302.44 | 283.78 | 293.11 | 150.81 | 122.16 | 136.48 | 387.95 | 320.48 | 354.44 | |
| α-$UH_3$ | Cubic | 0 | 126.43 | 126.43 | 126.43 | 86.13 | 78.34 | 82.23 | 210.57 | 194.78 | 202.74 | |
| | *Pm-3n* | 10 | 164.57 | 164.57 | 164.57 | 89.36 | 73.12 | 81.24 | 226.99 | 191.06 | 209.28 | [This] |
| | 223 | 20 | 198.91 | 198.91 | 198.91 | 85.74 | 49.96 | 67.85 | 224.90 | 138.29 | 182.76 | |
| $UH_5$ | Hexagonal | 10 | 175.25 | 174.46 | 174.86 | 99.07 | 96.18 | 97.62 | 250.09 | 243.74 | 246.92 | |
| | $P6_3mc$ | 20 | 213.41 | 212.16 | 212.79 | 109.55 | 107.66 | 108.61 | 280.63 | 276.25 | 278.44 | [This] |
| | 186 | 30 | 248.29 | 246.28 | 247.28 | 117.99 | 116.20 | 117.10 | 305.57 | 301.23 | 303.40 | |
| $UH_6$ | Hexagonal | 10 | 186.08 | 183.69 | 184.88 | 126.50 | 124.52 | 125.51 | 309.39 | 304.72 | 307.05 | |
| | $P6_3/mmc$ | 20 | 223.59 | 221.48 | 222.53 | 139.49 | 137.49 | 138.49 | 346.43 | 341.75 | 344.09 | [This] |
| | 194 | 30 | 258.95 | 257.28 | 258.12 | 152.24 | 150.18 | 151.21 | 381.88 | 377.15 | 379.51 | |
| $UH_7$ | Hexagonal | 0 | 155.90 | 155.23 | 155.56 | 129.10 | 128.23 | 128.66 | 303.52 | 301.63 | 302.57 | |
| | $P6_3/mmc$ | 10 | 194.14 | 193.64 | 193.89 | 148.18 | 147.50 | 147.84 | 354.37 | 352.90 | 353.64 | |
| | 194 | 20 | 232.26 | 231.86 | 232.06 | 165.48 | 165.05 | 165.27 | 401.17 | 400.19 | 400.68 | [This] |
| | | 30 | 266.57 | 266.41 | 266.49 | 179.07 | 178.87 | 178.97 | 438.93 | 438.48 | 438.71 | |
| | | 40 | 302.29 | 302.04 | 302.16 | 194.02 | 193.80 | 193.91 | 479.48 | 478.95 | 479.22 | |
| $UH_8$ | Cubic | 0 | 160.12 | 160.12 | 160.12 | 144.17 | 143.07 | 143.62 | 332.66 | 330.71 | 331.68 | |
| | *Fm-3m* | 30 | 270.59 | 270.59 | 270.59 | 210.31 | 205.71 | 208.01 | 501.10 | 492.37 | 496.74 | |
| | 225 | 40 | 312.76 | 312.76 | 312.76 | 230.21 | 224.24 | 227.22 | 554.57 | 542.95 | 548.78 | |
| | | 50 | 339.54 | 339.54 | 339.54 | 245.55 | 239.54 | 242.54 | 593.57 | 581.79 | 587.69 | [This] |
| | | 60 | 372.93 | 372.93 | 372.93 | 262.86 | 255.98 | 259.42 | 638.55 | 624.96 | 631.77 | |

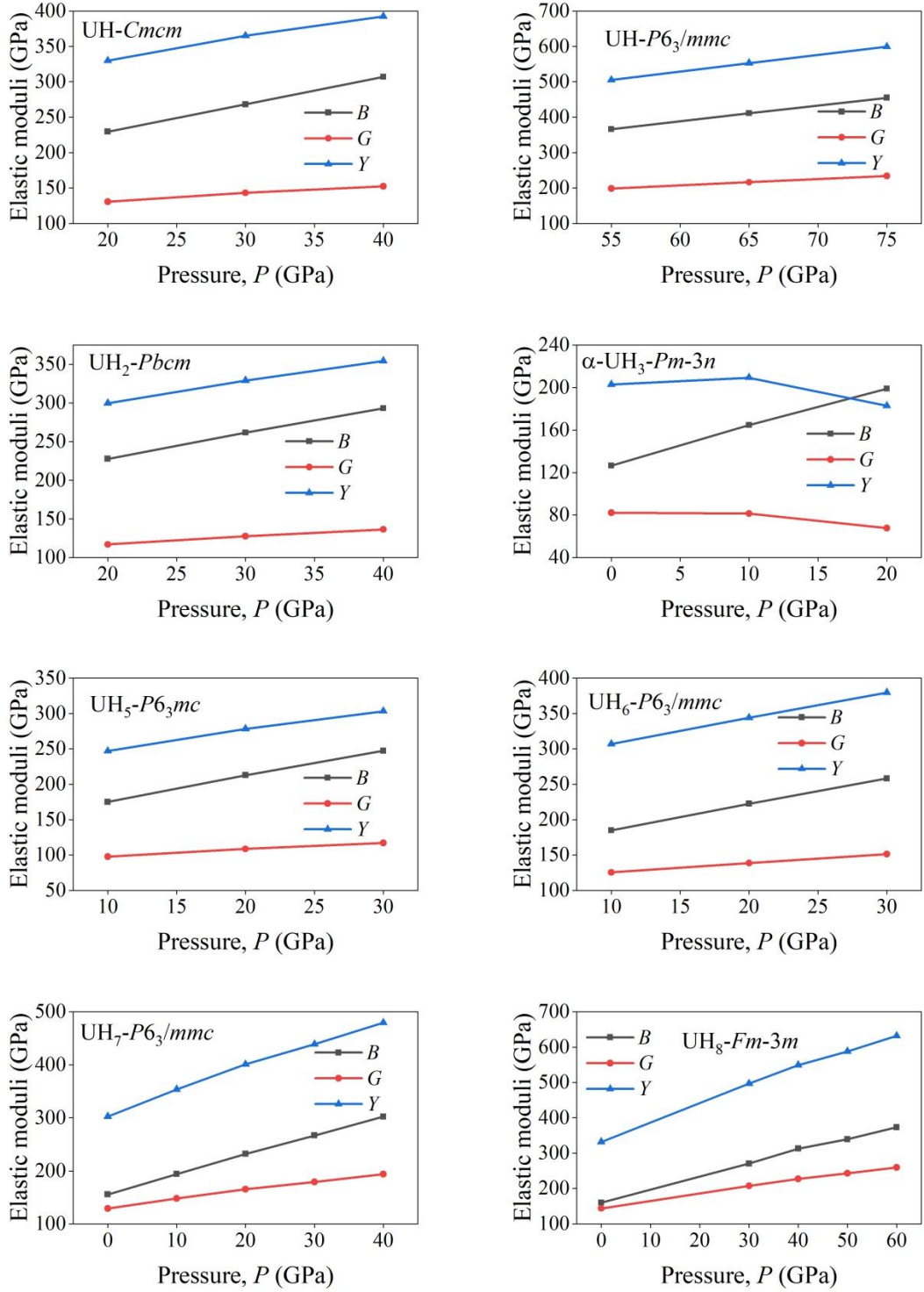

**Figure 5**: Elastic moduli of UH$_x$ ($x$ = 1, 2, 3, 5, 6, 7, 8) compounds under different pressures.

**Table 5**. Minimal and maximal values of the elastic moduli ($Y$ and $G$), linear compressibility ($\beta$), and Poisson's ratio ($v$) for UH$_x$ ($x$ = 1, 2, 3, 5, 6, 7, 8) compounds under different pressures.

| Compounds | Crystal System Space Group Space Group No. | Pressure, $P$ (GPa) | Young, $Y$ (GPa) | | Shear, $G$ (GPa) | | Linear compressibility (TPa$^{-1}$) | | Poisson's ratio | | Ref. |
|---|---|---|---|---|---|---|---|---|---|---|---|
| | | | $Y_{min}$ | $Y_{max}$ | $G_{min}$ | $G_{max}$ | $\beta_{min}$ | $\beta_{max}$ | $v_{min}$ | $v_{max}$ | |
| UH | Orthorhombic | 20 | 245.17 | 495.31 | 87.15 | 201.28 | 1.108 | 2.153 | 0.095 | 0.468 | |
| | *Cmcm* | 30 | 288.72 | 555.08 | 92.19 | 222.68 | 0.989 | 1.839 | 0.081 | 0.498 | [This] |
| | 63 | 40 | 268.00 | 615.96 | 86.19 | 244.99 | 0.830 | 1.591 | 0.087 | 0.560 | |
| UH | Hexagonal | 55 | 373.93 | 787.9 | 123.25 | 320.18 | 0.871 | 0.990 | 0.110 | 0.521 | |
| | P6$_3$/mmc | 65 | 410.97 | 868.4 | 134.88 | 345.95 | 0.788 | 0.856 | 0.114 | 0.528 | [This] |
| | 194 | 75 | 449.98 | 938.13 | 147.83 | 368.56 | 0.718 | 0.762 | 0.122 | 0.526 | |
| UH$_2$ | Orthorhombic | 20 | 219.7 | 428.66 | 73.351 | 170.51 | 0.932 | 2.134 | 0.108 | 0.536 | |
| | *Pbcm* | 30 | 215.49 | 497.65 | 71.695 | 199.7 | 0.778 | 1.933 | 0.111 | 0.542 | [This] |
| | 57 | 40 | 210.68 | 559.2 | 68.013 | 224.88 | 0.660 | 1.799 | 0.099 | 0.581 | |
| α-UH$_3$ | Cubic | 0 | 163.48 | 273.25 | 63.636 | 119.87 | 2.636 | 2.636 | 0.093 | 0.428 | |
| | *Pm-3n* | 10 | 149.51 | 327.66 | 55.432 | 140.24 | 2.026 | 2.026 | 0.089 | 0.561 | [This] |
| | 223 | 20 | 96.928 | 384.29 | 34.159 | 163.11 | 1.676 | 1.676 | 0.055 | 0.745 | |
| UH$_5$ | Hexagonal | 10 | 230.94 | 285.88 | 80.409 | 109.25 | 1.780 | 2.173 | 0.153 | 0.436 | |
| | *P6$_3$mc* | 20 | 269.32 | 316.43 | 95.878 | 122.72 | 1.441 | 1.832 | 0.181 | 0.425 | [This] |
| | 186 | 30 | 288.6 | 344.89 | 109.14 | 135.31 | 1.217 | 1.626 | 0.198 | 0.424 | |
| UH$_6$ | Hexagonal | 10 | 289.45 | 401.54 | 114.87 | 142.87 | 1.428 | 2.008 | 0.154 | 0.263 | |
| | P6$_3$/mmc | 20 | 326.38 | 443.94 | 127.79 | 158.05 | 1.218 | 1.649 | 0.170 | 0.287 | [This] |
| | 194 | 30 | 361.25 | 483.51 | 140.41 | 173.07 | 1.085 | 1.401 | 0.180 | 0.306 | |
| | | 0 | 286.12 | 349.65 | 118.1 | 136.61 | 1.895 | 2.274 | 0.138 | 0.211 | |
| UH$_7$ | Hexagonal | 10 | 337.47 | 395.98 | 137.44 | 154.85 | 1.556 | 1.804 | 0.164 | 0.228 | |
| | P6$_3$/mmc | 20 | 387.44 | 439.54 | 156.84 | 172.08 | 1.320 | 1.497 | 0.185 | 0.235 | [This] |
| | 194 | 30 | 427.51 | 460.6 | 172.24 | 182.63 | 1.187 | 1.283 | 0.210 | 0.241 | |
| | | 40 | 466.94 | 506.3 | 186.85 | 197.78 | 1.036 | 1.138 | 0.219 | 0.249 | |
| UH$_8$ | Cubic | 0 | 313.77 | 359.84 | 133.7 | 159.87 | 2.082 | 2.082 | 0.113 | 0.212 | |
| | *Fm-3m* | 30 | 450.31 | 572.59 | 184.16 | 249.53 | 1.232 | 1.232 | 0.122 | 0.292 | |
| | 225 | 40 | 492.37 | 641.86 | 198.92 | 277.15 | 1.066 | 1.066 | 0.129 | 0.314 | |
| | | 50 | 528.8 | 684.72 | 213.15 | 294.15 | 0.982 | 0.982 | 0.134 | 0.315 | [This] |
| | | 60 | 566.11 | 740.42 | 226.99 | 316.66 | 0.894 | 0.894 | 0.137 | 0.325 | |

Maximal and minimal values of various elastic moduli and Poisson's ratio are presented in Table 5. These values were obtained from the ELATE [37] analysis of direction dependent elastic properties of UH$_x$ ($x$ = 1, 2, 3, 5, 6, 7, 8) compounds. The ratio between the maximum and minimum values is a measure of the direction dependent anisotropy in the atomic bonding within the crystals. From Table 5 it is seen that anisotropy in the modulus of rigidity and Poisson's ratio is higher than the anisotropy in the Young's modulus and compressibility.

**Table 6**. Calculated Poisson's ratio ($v$), Pugh's ratio ($G/B$), Grüneisen parameter ($\gamma$), and machinability index ($\mu_m$) of UH$_x$ ($x$ = 1, 2, 3, 5, 6, 7, 8) compounds under different pressures.

| Compounds | Crystal System Space Group Space Group No. | Pressure, $P$ (GPa) | $v$ | $G/B$ | $\gamma$ | $\mu_m$ | Ref. |
|---|---|---|---|---|---|---|---|
| UH | Orthorhombic | 20 | 0.26 | 0.57 | 1.55 | 1.14 | |
|  | *Cmcm* | 30 | 0.27 | 0.53 | 1.62 | 1.20 | [This] |
|  | 63 | 40 | 0.29 | 0.50 | 1.69 | 1.25 | |
| UH | Hexagonal | 55 | 0.27 | 0.54 | 1.60 | 2.97 | |
|  | *P6$_3$/mmc* | 65 | 0.28 | 0.53 | 1.63 | 3.05 | [This] |
|  | 194 | 75 | 0.28 | 0.51 | 1.66 | 3.08 | |
| UH$_2$ | Orthorhombic | 20 | 0.28 | 0.51 | 1.66 | 1.34 | |
|  | *Pbcm* | 30 | 0.29 | 0.49 | 1.71 | 1.31 | [This] |
|  | 57 | 40 | 0.30 | 0.47 | 1.76 | 1.30 | |
| α-UH$_3$ | Cubic | 0 | 0.23 | 0.65 | 1.42 | 1.99 | |
|  | *Pm-3n* | 10 | 0.29 | 0.49 | 1.70 | 2.97 | [This] |
|  | 223 | 20 | 0.35 | 0.34 | 2.11 | 5.82 | |
| UH$_5$ | Hexagonal | 10 | 0.26 | 0.56 | 1.57 | 1.67 | |
|  | *P6$_3$mc* | 20 | 0.28 | 0.51 | 1.67 | 1.94 | [This] |
|  | 186 | 30 | 0.30 | 0.47 | 1.75 | 2.21 | |
| UH$_6$ | Hexagonal | 10 | 0.22 | 0.68 | 1.38 | 1.51 | |
|  | *P6$_3$/mmc* | 20 | 0.24 | 0.62 | 1.46 | 1.66 | [This] |
|  | 194 | 30 | 0.25 | 0.59 | 1.52 | 1.77 | |
| UH$_7$ | Hexagonal | 0 | 0.18 | 0.83 | 1.20 | 1.18 | |
|  | *P6$_3$/mmc* | 10 | 0.20 | 0.76 | 1.27 | 1.28 | |
|  | 194 | 20 | 0.21 | 0.71 | 1.33 | 1.38 | [This] |
|  |  | 30 | 0.23 | 0.67 | 1.39 | 1.46 | |
|  |  | 40 | 0.24 | 0.64 | 1.43 | 1.53 | |
| UH$_8$ | Cubic | 0 | 0.15 | 0.90 | 1.13 | 1.20 | |
|  | *Fm-3m* | 30 | 0.19 | 0.77 | 1.26 | 1.47 | |
|  | 225 | 40 | 0.21 | 0.73 | 1.32 | 1.57 | |
|  |  | 50 | 0.21 | 0.71 | 1.33 | 1.59 | [This] |
|  |  | 60 | 0.22 | 0.70 | 1.36 | 1.64 | |

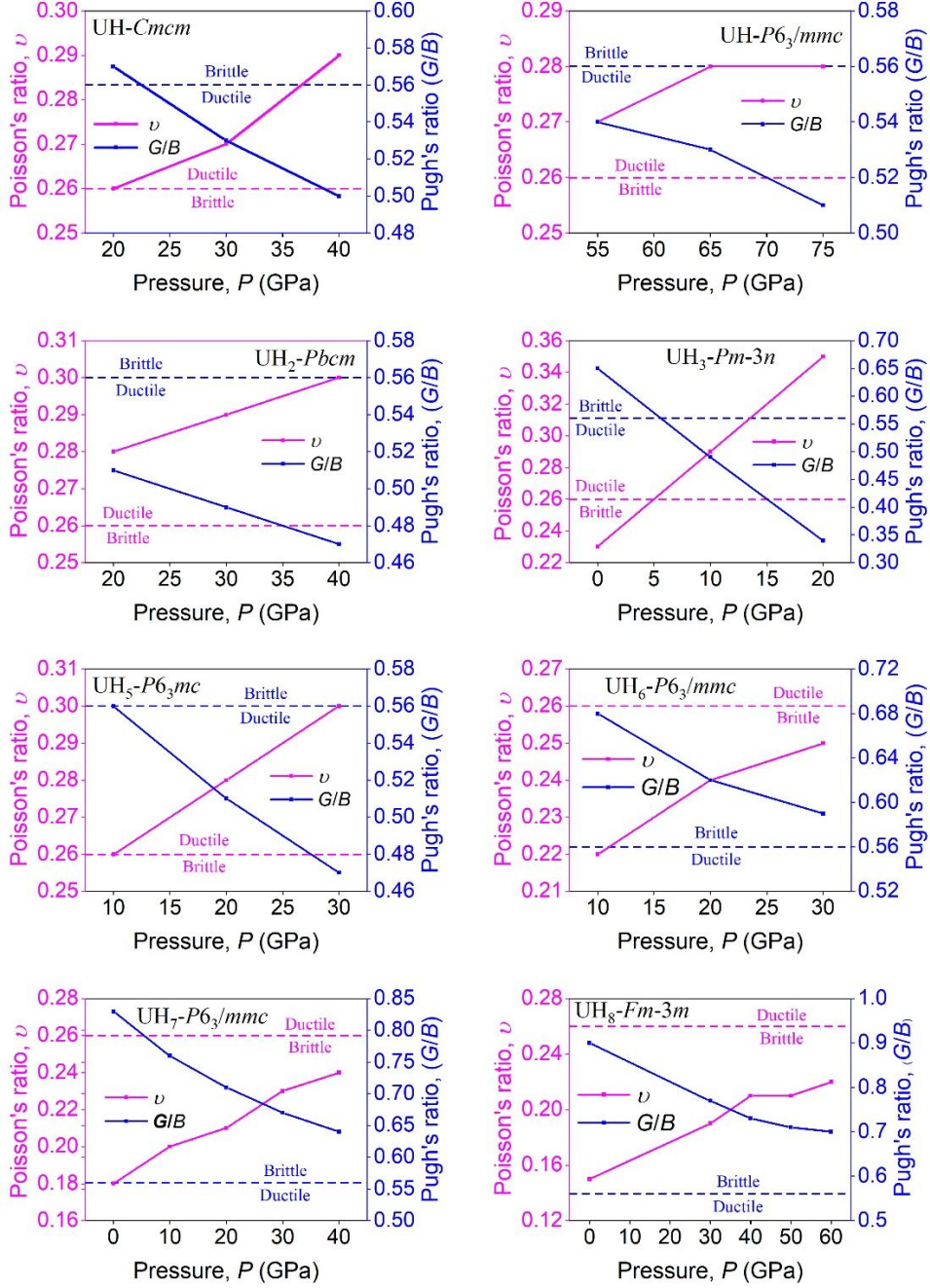

**Figure 6**: Poisson's and Pugh's ratio of UH$_x$ ($x$ = 1, 2, 3, 5, 6, 7, 8) compounds under different pressures.

The Poisson's ratio $v$ is calculated using the equation: $v = (3B - 2G)/(6B + 2G)$. The Pugh's ratio $G/B$ and machinability index $\mu_m$ (= $B/C_{44}$) are also computed and listed in Table 6. These three parameters are useful elastic/mechanical indicators. For example Poisson's ratio and Pugh's ratio can differentiate brittleness and ductility in solids. According to Frantsevich [38], if $v > 0.26$ the solid should exhibit ductile nature otherwise it should be brittle. From Table 6 and Fig. 6 we see that the values of $v < 0.26$ for α-UH$_3$-*Pm-3n* at 0 GPa, for UH$_6$-*P6$_3$/mmc* in the pressure range 10-30 GPa, for UH$_7$-*P6$_3$/mmc* in the pressure range 0-40 GPa and from UH$_8$ in the pressure range 0-60 GPa the compounds are predicted to be brittle. On the other hand, $v > 0.26$ for UH-*Cmcm* at 30 and 40 GPa, for UH-*P6$_3$/mmc* within the pressure range 55-75 GPa, for UH$_2$-*Pbcm* within the pressure range 20-30 GPa, for α-UH$_3$-*Pm-3n* at 10 and 20 GPa, for UH$_6$-*P6$_3$mc* in the pressure range 10-30 GPa, ductility is expected. The Pugh's ratio (*G/B*) [39] is another parameter used to judge ductile and brittle nature of a solid [39]. $G/B > 0.57$ indicates the brittle nature and otherwise ductility. From Table 6 and Fig. 6 it is observed that the Pugh's ratio gives the same information about brittleness/ductility as the Poisson's ratio.

The machinability index determines the ease with which a solid can be shaped into desired form. It is also an indicator of dry lubricity of solids. High machinability index implies ease in the cutting and low level of frictional loss. For all the compounds under study, the machinability index increases systematically with increasing pressure. For the compound α-UH$_3$ in the cubic phase, $\mu_m$ is very high (5.82) at 20 GPa, suggesting that this compound becomes extremely machinable and should have excellent dry lubricity at high pressures.

Grüneisen parameter is related to the anharmonicity in the lattice dynamics. High value of Grüneisen parameter is indicative of strong electron-phonon interaction in superconductors [40]. This is because anharmonic and soft phonon modes contribute strongly to the electron-phonon coupling constant. The pressure dependent values of the Grüneisen parameter, calculated using the Poisson's ratio, are tabulated in Table 6. The obtained values of $\gamma$ are typical [41]. The highest value of the Grüneisen parameter (2.11) among all the compounds was found for α-UH$_3$ in the cubic phase at 20 GPa.

*3.3 Elastic anisotropy*

Study of the elastic anisotropy is important to understand the direction dependent bonding characteristics and mechanical properties of crystalline solids. The shear anisotropy factors [42] obtained from the elastic constants are given in Table 7. These factors $A_1$, $A_2$, and $A_3$ are one for an isotropic crystal, while any value except unity is a measure of the degree of elastic anisotropy possessed by the crystal. The factors $A_1$, $A_2$, and $A_3$ are computed from the expressions given below:

$$A_1 = \frac{4C_{44}}{C_{11} + C_{33} - 2C_{13}}, A_2 = \frac{4C_{55}}{C_{22} + C_{33} - 2C_{23}}, A_3 = \frac{4C_{66}}{C_{11} + C_{22} - 2C_{12}} \quad (6)$$

Furthermore, $A^B$ and $A^G$ are the percentage anisotropies in compressibility and shear, respectively, and $A^U$ is the universal anisotropy index. The zero values of $A^B$, $A^G$ and $A^U$ represent elastic isotropy of a crystal and non-zero values represent anisotropy [43, 44]. The computed values of $A^U$, $A^B$, and $A^G$ are also listed in Table 7. These anisotropy indices are calculated using the following equations:

$$A^B = \frac{B_V - B_R}{B_V + B_R}, A^G = \frac{G_V - G_R}{G_V + G_R}, and \; A^U = 5\frac{G_V}{G_R} + \frac{B_V}{B_R} - 6 \geq 0 \quad (7)$$

Table 7 (and Fig. 7) shows clearly that the $UH_x$ ($x$ = 1, 2, 3, 5, 6, 7, 8) compounds are elastically anisotropic. These anisotropies are clear indications that atomic bonding strengths in different directions in the crystals are different. Among the compounds of interest, hexagonal $UH_7$-$P6_3/mmc$ has the lowest anisotropy because $A_1 = A_2 = A_3 = 1$ and $A^U$, $A^B$ and $A^G$ values are very close to zero in the pressure range 30 – 40 GPa. Overall the orthorhombic UH-$Cmcm$ has the highest level of anisotropy. The pressure dependent variations in various anisotropy indices are nonmonotonic, indicating that changes in the bonding characters are different in different crystal directions when uniform hydrostatic pressure is applied. Fig. 7 shows these variations clearly.

**Table 7**. The shear anisotropy factors $A_1$, $A_2$, $A_3$, and $A^B$ (in %), $A^G$ (in %), $A^U$ of $UH_x$ ($x$ = 1, 2, 3, 5, 6, 7, 8) compounds under different pressures.

| Compounds | Crystal System Space Group Space Group No. | Pressure, $P$ (GPa) | $A_1$ | $A_2$ | $A_3$ | $A^B$ | $A^G$ | $A^U$ | Ref. |
|---|---|---|---|---|---|---|---|---|---|
| UH | Orthorhombic | 20 | 1.13 | 0.57 | 0.59 | 2.47 | 6.36 | 0.73 | |
| | $Cmcm$ | 30 | 1.10 | 0.56 | 0.56 | 2.29 | 7.06 | 0.81 | [This] |
| | 63 | 40 | 1.07 | 0.55 | 0.48 | 2.11 | 8.92 | 1.02 | |
| UH | Hexagonal | 55 | 0.37 | 0.37 | 1.00 | 0.13 | 8.64 | 0.95 | |
| | $P6_3/mmc$ | 65 | 0.38 | 0.38 | 1.00 | 0.05 | 8.38 | 0.92 | [This] |
| | 194 | 75 | 0.39 | 0.39 | 1.00 | 0.03 | 7.88 | 0.86 | |
| $UH_2$ | Orthorhombic | 20 | 1.09 | 0.48 | 0.50 | 2.30 | 7.10 | 0.81 | |
| | $Pbcm$ | 30 | 1.15 | 0.41 | 0.52 | 2.71 | 8.85 | 1.03 | [This] |

|  |  | 57 | 40 | 1.19 | 0.37 | 0.56 | 3.18 | 10.50 | 1.24 |  |
| --- | --- | --- | --- | --- | --- | --- | --- | --- | --- | --- |
| α-UH$_3$ | Cubic | 0 | 0.53 | 0.53 | 0.53 | 0.00 | 4.74 | 0.50 |  |
|  | *Pm-3n* | 10 | 0.40 | 0.40 | 0.40 | 0.00 | 9.99 | 1.11 |  |
|  | 223 | 20 | 0.21 | 0.21 | 0.21 | 0.00 | 26.37 | 3.58 |  |
| UH$_5$ | Hexagonal | 10 | 0.92 | 0.92 | 1.00 | 0.22 | 1.48 | 0.16 |  |
|  | *P6$_3$mc* | 20 | 0.87 | 0.87 | 1.00 | 0.29 | 0.87 | 0.09 | [This] |
|  | 186 | 30 | 0.81 | 0.81 | 1.00 | 0.40 | 0.76 | 0.09 |  |
| UH$_6$ | Hexagonal | 10 | 0.83 | 0.83 | 1.00 | 0.65 | 0.79 | 0.09 |  |
|  | *P6$_3$/mmc* | 20 | 0.83 | 0.83 | 1.00 | 0.47 | 0.72 | 0.08 | [This] |
|  | 194 | 30 | 0.82 | 0.82 | 1.00 | 0.32 | 0.68 | 0.08 |  |
| UH$_7$ | Hexagonal | 0 | 0.96 | 0.96 | 1.00 | 0.21 | 0.34 | 0.04 |  |
|  | *P6$_3$/mmc* | 10 | 0.97 | 0.97 | 1.00 | 0.13 | 0.23 | 0.03 |  |
|  | 194 | 20 | 0.97 | 0.97 | 1.00 | 0.09 | 0.13 | 0.01 | [This] |
|  |  | 30 | 1.00 | 1.00 | 1.00 | 0.03 | 0.06 | 0.01 |  |
|  |  | 40 | 1.00 | 1.00 | 1.00 | 0.04 | 0.06 | 0.01 |  |
| UH$_8$ | Cubic | 0 | 0.84 | 0.84 | 0.84 | 0.00 | 0.38 | 0.04 |  |
|  | *Fm-3m* | 10 | 0.74 | 0.74 | 0.74 | 0.00 | 1.10 | 0.11 |  |
|  | 225 | 40 | 0.72 | 0.72 | 0.72 | 0.00 | 1.31 | 0.13 |  |
|  |  | 50 | 0.72 | 0.72 | 0.72 | 0.00 | 1.24 | 0.13 | [This] |
|  |  | 60 | 0.72 | 0.72 | 0.72 | 0.00 | 1.32 | 0.13 |  |

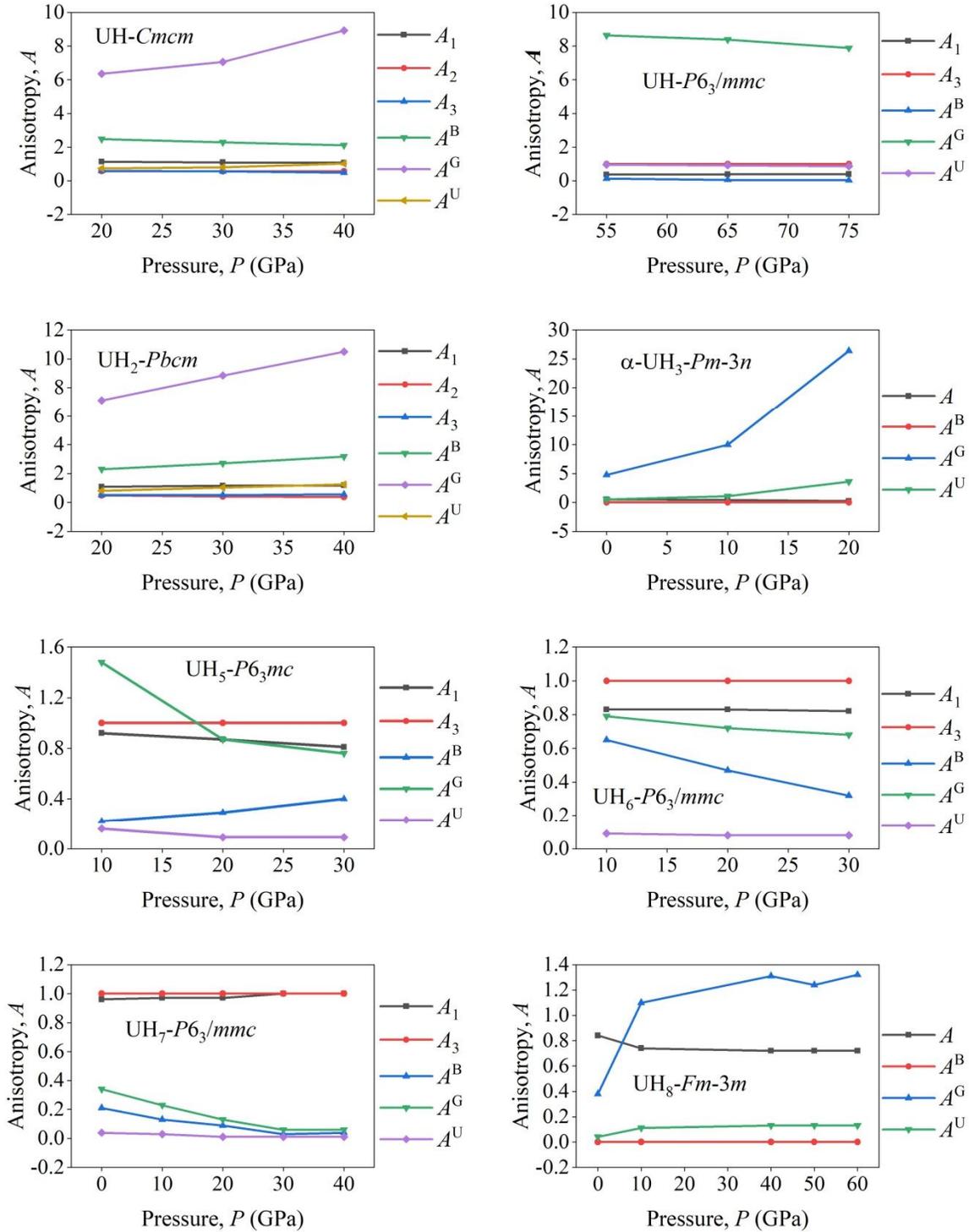

**Figure 7**: Anisotropy indices of UH$_x$ ($x$ = 1, 2, 3, 5, 6, 7, 8) compounds under different pressures.

## 3.4 Acoustic velocities

The sound velocities in a crystal are useful thermophysical parameter. These velocities are closely related to the crystal stiffness and crystal density, and determine the Debye temperature and thermal conductivity to a large extent [45]. Crystalline solids support both longitudinal and transverse modes of propagation of acoustic disturbances. The phonon thermal conductivity of solids increases with increasing acoustic velocities.

**Table 8**. Calculated density $\rho$ (gm/cm$^3$), transverse sound velocities $v_t$ (km/s), longitudinal sound velocities $v_l$ (km/s), average sound velocities $v_m$ (km/s), Debye temperature $\theta_D$ (K) and melting temperature $T_m$ (K) of UH$_x$ ($x$ = 1, 2, 3, 5, 6, 7, 8) compounds under different pressures.

| Compounds | Crystal System Space Group Space Group No. | Pressure, $P$ (GPa) | $\rho$ | $v_t$ | $v_l$ | $v_m$ | $\theta_D$ | $T_m$ | Ref |
|---|---|---|---|---|---|---|---|---|---|
| UH | Orthorhombic | 20 | 18.68 | 2.6464 | 4.6512 | 2.9417 | 398.41 | 2218.34 | |
| | *Cmcm* | 30 | 19.46 | 2.7138 | 4.8583 | 3.0212 | 414.80 | 2495.30 | [This] |
| | 63 | 40 | 20.17 | 2.7499 | 5.0303 | 3.0666 | 426.06 | 2780.27 | |
| UH | Hexagonal | 55 | 21.98 | 3.0080 | 5.3607 | 3.3475 | 478.61 | 3804.63 | |
| | *P6$_3$/mmc* | 65 | 22.55 | 3.0996 | 5.5716 | 3.4518 | 497.80 | 4150.13 | [This] |
| | 194 | 75 | 23.08 | 3.1835 | 5.7652 | 3.5474 | 515.52 | 4475.74 | |
| UH$_2$ | Orthorhombic | 20 | 16.13 | 2.6934 | 4.8774 | 3.0011 | 442.45 | 2083.43 | |
| | *Pbcm* | 30 | 16.82 | 2.7535 | 5.0656 | 3.0719 | 459.27 | 2294.97 | [This] |
| | 57 | 40 | 17.46 | 2.7959 | 5.2165 | 3.1224 | 472.63 | 2482.08 | |
| α-UH$_3$ | Cubic | 0 | 12.06 | 2.6108 | 4.4237 | 2.8928 | 425.47 | 1642.15 | |
| | *Pm-3n* | 10 | 12.92 | 2.5073 | 4.5953 | 2.7964 | 420.84 | 1936.02 | [This] |
| | 223 | 20 | 13.65 | 2.2292 | 4.6037 | 2.5052 | 384.00 | 2227.75 | |
| UH$_5$ | Hexagonal | 10 | 11.20 | 2.9520 | 5.2180 | 3.2830 | 537.78 | 1757.75 | |
| | *P6$_3$mc* | 20 | 11.79 | 3.0351 | 5.5074 | 3.3825 | 563.59 | 2010.43 | [This] |
| | 186 | 30 | 12.31 | 3.0839 | 5.7241 | 3.4427 | 581.98 | 2243.46 | |
| UH$_6$ | Hexagonal | 10 | 10.63 | 3.4369 | 5.7576 | 3.8040 | 643.61 | 2008.54 | |
| | *P6$_3$/mmc* | 20 | 11.17 | 3.5217 | 6.0386 | 3.9062 | 671.95 | 2263.19 | [This] |
| | 194 | 30 | 11.64 | 3.6043 | 6.2847 | 4.0038 | 698.32 | 2506.96 | |
| | | 0 | 9.86 | 3.6115 | 5.7585 | 3.9769 | 685.36 | 1844.10 | |
| UH$_7$ | Hexagonal | 10 | 10.44 | 3.7624 | 6.1187 | 4.1518 | 729.24 | 2126.23 | |
| | *P6$_3$/mmc* | 20 | 10.94 | 3.8868 | 6.4309 | 4.2967 | 766.43 | 2404.09 | |
| | 194 | 30 | 11.38 | 3.9649 | 6.6609 | 4.3895 | 793.47 | 2627.19 | |
| | | 40 | 11.79 | 4.0550 | 6.8955 | 4.4944 | 822.02 | 2880.52 | |
| UH$_8$ | | 0 | 9.65 | 3.8585 | 6.0374 | 4.2399 | 753.26 | 2033.72 | |
| | Cubic | 10 | 11.11 | 4.3276 | 7.0237 | 4.7745 | 889.05 | 3068.84 | |

|   | Fm-3m | 40 | 11.49 | 4.4468 | 7.3200 | 4.9132 | 925.30 | 3424.33 | [This] |
|   | 225 | 50 | 11.85 | 4.5236 | 7.4786 | 5.0002 | 951.47 | 3646.83 |   |
|   |   | 60 | 12.19 | 4.6131 | 7.6789 | 5.1026 | 980.08 | 3932.16 |   |

The calculated values of sound velocities ($v_t$, $v_l$ and $v_m$) under pressure of UH$_x$ compounds are listed in Table 8. The following equations were used to determine the sound velocities [46-48]:

$$v_t = \sqrt{\frac{G}{\rho}}, \; v_l = \sqrt{\frac{3B+4G}{3\rho}} \text{ and } v_m = \left[\frac{1}{3}\left(\frac{2}{v_t^3}+\frac{1}{v_l^3}\right)\right]^{-\frac{1}{3}} \tag{8}$$

In the above equations $v_t$, $v_l$ and $v_m$ signify the transverse, longitudinal, and the mean sound velocities, respectively. The computed values of sound velocities at different pressures are enlisted in Table 8. The pressure dependent variations in the sound velocities are also illustrated in Fig. 8 for all the compounds under study. In general, the sound velocities increase with increasing pressure due to the increase in the crystal stiffness. This is seen in all the UH$_x$ compounds, except the cubic α-UH$_3$-$Pm$-$3n$ compound. This particular compound also shows unconventional variation of the elastic moduli with pressure variation.

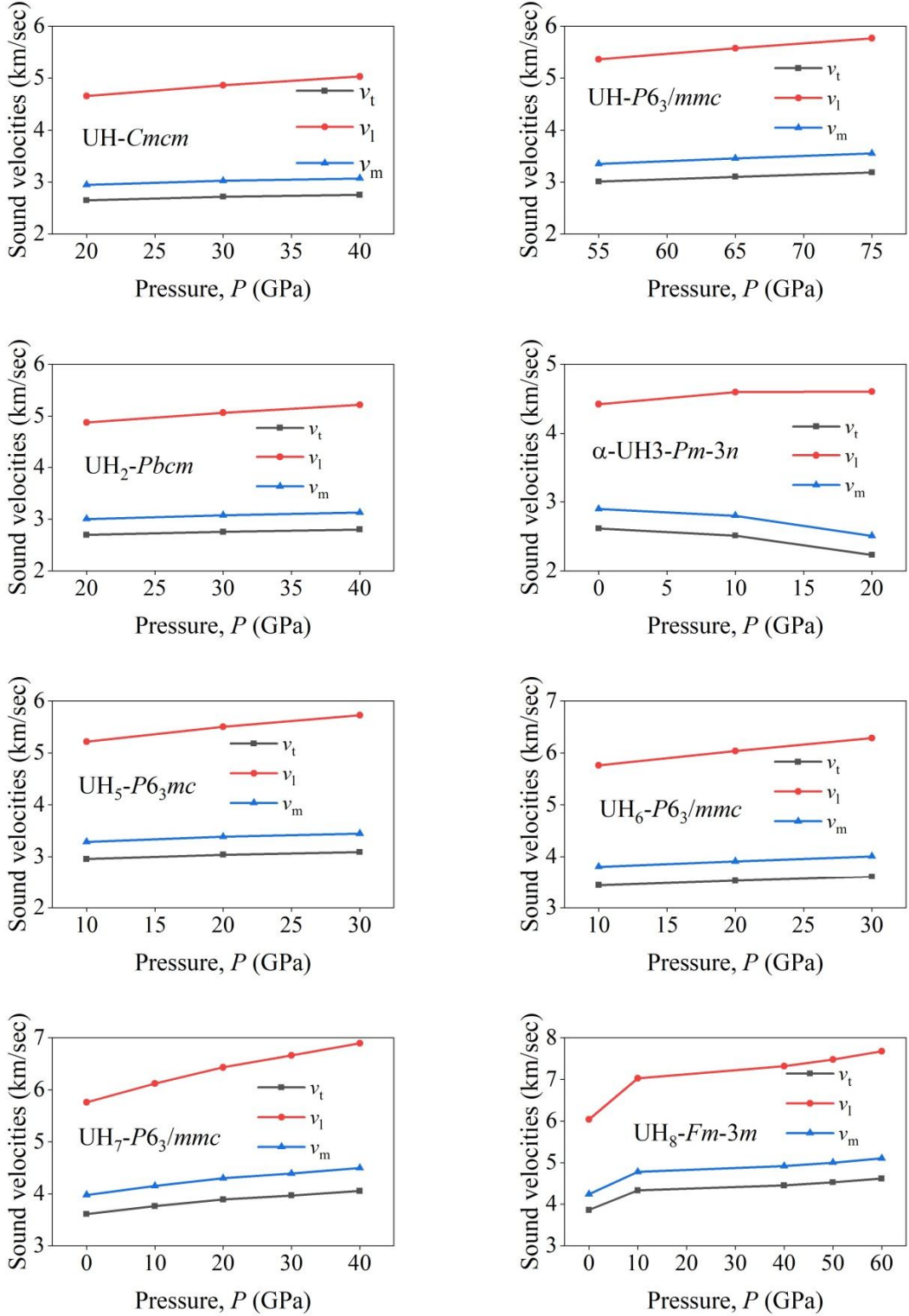

**Figure 8**: Sound velocities in UH$_x$ ($x$ = 1, 2, 3, 5, 6, 7, 8) compounds under different pressures.

*3.5 Hardness of UH$_x$ compounds*

Hardness of a solid is used to assess both elastic and plastic behaviors of the material under mechanical stress. This particular parameter determines the average bonding strength and stiffness of a solid. A variety of theoretical methodologies are available to compute the hardness of a solid. In this study, we have calculated the hardness of the UH$_x$ ($x$ = 1, 2, 3, 5, 6, 7, 8) compounds at different pressures using the formalisms developed by Teter et al. [49], Tian et al. [50], Chen et al. [51], microhardness [52], and Efim Mazhnik [53]. The calculated values of hardness are presented in Table 9. The pressure dependent hardness values are also illustrated in Fig. 9 showing that different formalisms result in different values of hardness. Considering the highest values of the hardness, the calculated values according to Teter et al. are arranged in the following order: UH-*P*6$_3$/*mmc* > UH$_8$-*Fm-3m* > UH-*Cmcm* > UH$_7$-*P*6$_3$/*mmc* > UH$_6$-*P*6$_3$/*mmc* > UH$_2$-*Pbcm* > UH$_5$-*P6$_3$mc* > α-UH$_3$-*Pm-3n*. According to Tian et al., the hardnesses are arranged in the following order: UH$_8$-*Fm-3m* > UH$_7$-*P*6$_3$/*mmc* > UH-*P*6$_3$/*mmc* > UH$_6$-*P*6$_3$/*mmc* > UH-*Cmcm* > α-UH$_3$-*Pm-3n* > UH$_2$-*Pbcm* > UH$_5$- *P6$_3$mc*. According to Chen et al., the hardnesses are arranged in the following order: UH$_8$-*Fm-3m* > UH$_7$-*P*6$_3$/*mmc* > UH-*P*6$_3$/*mmc* > UH$_6$-*P*6$_3$/*mmc* > UH-*Cmcm* > α-UH$_3$- *Pm-3n* > UH$_5$-*P6$_3$mc* > UH$_2$-*Pbcm*. According to microhardness, the hardnesses values arranged in the following order: UH$_8$-*Fm-3m* > UH- *P*6$_3$/*mmc* > UH$_7$-*P*6$_3$/*mmc* > UH$_6$-*P*6$_3$/*mmc* > UH-*Cmcm* > UH$_2$-*Pbcm* > UH$_5$-*P6$_3$mc* > α-UH$_3$-*Pm-3n*. According to Efim-Mazhnik, the hardnesses values are arranged in the following order: UH$_8$-*Fm-3m* > UH-*P*6$_3$/*mmc* > UH$_7$- *P*6$_3$/*mmc* > UH-*Cmcm* > UH$_2$-*Pbcm* > UH$_6$-*P*6$_3$/*mmc* > UH$_5$-*P6$_3$mc* > α-UH$_3$-*Pm-3n*. It is interesting to note that the microhardness value of the cubic UH$_8$-*Fm-3m* is extremely high at high pressures and this compound can be classified as a superhard solid. Overall, among all the binary uranium hydrides considered, the cubic α-UH$_3$-*Pm-3n* phase seems to be the least hard. It should be stressed here that comparison among the hardness values in this section is rough measures since the pressures are different for the different structures. Nevertheless, the comparisons have been made with extrapolations and therefore, should result in fairly reliable conclusions.

**Table 9**. Hardness of UH$_x$ ($x$ = 1, 2, 3, 5, 6, 7, 8) compounds under different pressures.

| Compounds | Crystal System Space Group Space Group No. | Pressure, $P$ (GPa) | Hardness $H$ (GPa) | | | | | Ref. |
|---|---|---|---|---|---|---|---|---|
| | | | $H_{\text{Teter}}$ | $H_{\text{Tian}}$ | $H_{\text{Chen}}$ | $H_{\text{micro}}$ | $H_{\text{Efim Mazhnik}}$ | |
| UH | Orthorhombic | 20 | 19.75 | 13.41 | 14.92 | 20.88 | 15.78 | |
| | *Cmcm* | 30 | 21.64 | 13.11 | 14.54 | 21.67 | 18.01 | [This] |
| | 63 | 40 | 23.03 | 12.40 | 13.70 | 21.67 | 20.04 | |
| UH | Hexagonal | 55 | 30.03 | 16.88 | 18.63 | 30.46 | 24.74 | |
| | *P6$_3$/mmc* | 65 | 32.72 | 17.23 | 18.97 | 32.37 | 27.47 | [This] |
| | 194 | 75 | 35.32 | 17.57 | 19.31 | 34.20 | 30.12 | |
| UH$_2$ | Orthorhombic | 20 | 17.67 | 10.76 | 11.88 | 17.11 | 15.07 | |
| | *Pbcm* | 30 | 19.26 | 10.65 | 11.72 | 17.83 | 16.94 | [This] |
| | 57 | 40 | 20.61 | 10.49 | 11.51 | 18.34 | 18.57 | |
| α-UH$_3$ | Cubic | 0 | 12.42 | 11.58 | 12.95 | 14.65 | 9.42 | |
| | *Pm-3n* | 10 | 12.27 | 7.87 | 8.47 | 11.48 | 10.71 | [This] |
| | 223 | 20 | 10.25 | 4.17 | 3.70 | 6.93 | 10.26 | |
| UH$_5$ | Hexagonal | 10 | 14.74 | 10.61 | 11.75 | 15.32 | 11.93 | |
| | *P6$_3$mc* | 20 | 16.40 | 10.12 | 11.13 | 15.79 | 14.04 | |
| | 186 | 30 | 17.68 | 9.63 | 10.53 | 15.96 | 15.79 | |
| UH$_6$ | Hexagonal | 10 | 18.95 | 16.57 | 18.48 | 23.16 | 14.53 | |
| | *P6$_3$/mmc* | 20 | 20.91 | 15.77 | 17.55 | 23.79 | 15.97 | [This] |
| | 194 | 30 | 22.83 | 15.44 | 17.15 | 24.70 | 17.93 | |
| UH$_7$ | Hexagonal | 0 | 19.43 | 22.10 | 24.45 | 27.81 | 19.13 | |
| | *P6$_3$/mmc* | 10 | 22.32 | 21.81 | 24.08 | 29.96 | 19.39 | |
| | 194 | 20 | 24.96 | 21.49 | 23.68 | 31.71 | 19.84 | |
| | | 30 | 27.02 | 20.98 | 23.10 | 32.74 | 20.63 | |
| | | 40 | 29.28 | 20.87 | 22.93 | 34.17 | 22.22 | |
| UH$_8$ | Cubic | 0 | 21.69 | 26.69 | 29.19 | 33.06 | 23.69 | |
| | *Fm-3m* | 10 | 31.41 | 28.09 | 30.38 | 42.43 | 27.62 | |
| | 225 | 40 | 34.31 | 27.67 | 29.90 | 44.30 | 27.88 | |
| | | 50 | 36.62 | 28.32 | 30.51 | 46.64 | 29.20 | [This] |
| | | 60 | 39.17 | 28.64 | 30.79 | 48.83 | 30.50 | |

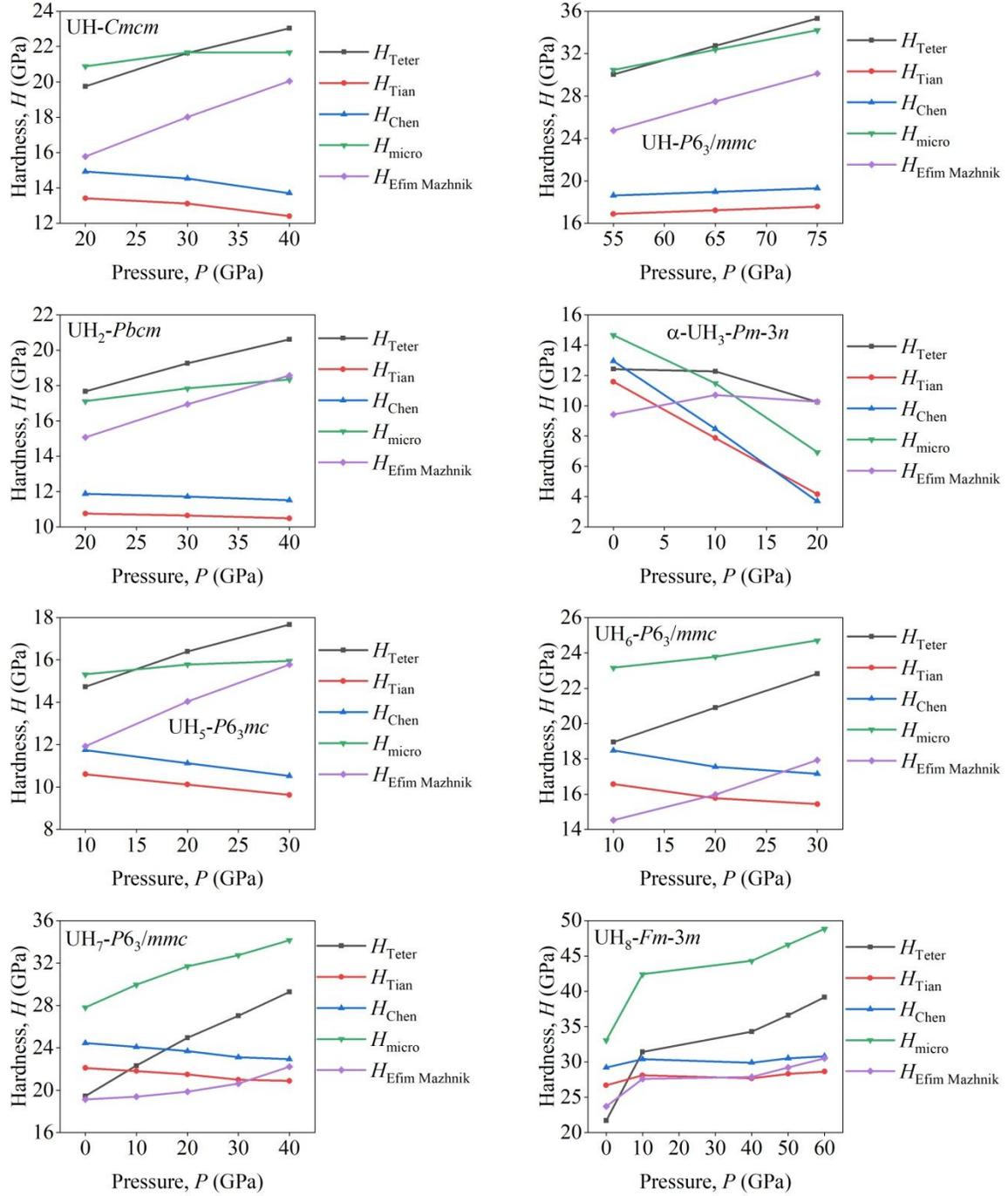

**Figure 9**: Hardness of UH$_x$ ($x$ = 1, 2, 3, 5, 6, 7, 8) compounds under different pressures.

## 3.6 Thermophysical properties of $UH_x$ compounds

*Debye temperature*: The Debye temperature $\theta_D$ of a system is closely connected to many major physical properties of solids such as the phonon specific heat, melting temperature, thermal conductivity, hardness, elastic constants, bonding strength, and sound velocity. The knowledge of $\theta_D$ provides information about the electron-phonon coupling and Cooper pairing mechanism of superconductivity. The Debye temperature calculated from the elastic constants is considered to be similar to that acquired from the temperature dependent specific heat measurements. Using the average sound velocity, the Debye temperature can be calculated by using the Anderson method [54] as follows:

$$\theta_D = \frac{h}{k_B}\left[\left(\frac{3n}{4\pi}\right)\frac{N_A \rho}{M}\right]^{1/3} v_m \qquad (9)$$

where, $h$ is Planck's constant, $k_B$ is the Boltzmann's constant, $\rho$ is the density, $N_A$ is the Avogadro number, $M$ is the molecular mass, and $v_m$ is the average sound velocity. This particular straightforward approach gives very reliable values of the Debye temperature of solids belonging to diverse categories (insulators, metals, topological compounds, magnetic materials, and superconductors) [55-59]. The computed results of the binary uranium hydrides are disclosed in Table 8. $UH_8$ in the cubic phase has the highest Debye temperature. The hardness of this compound is also the highest. It is worth noticing that, among all the hydrides under investigation, the predicted superconducting transition temperature of this particular one is the highest. We believe that high predicted $T_c$ in this compound is partly due to its very high Debye temperature. In conventional phonon mediated superconductors, the superconducting transition temperature is directly proportional to the Debye temperature. The pressure induced increased in the Debye temperature suggests that for superconducting compositions, the $T_c$ should increase with increasing pressure provided that the electron-phonon coupling constant does not decrease significantly due to the increase in the pressure.

*Minimum thermal conductivity*: At temperatures above $\theta_D$ the thermal conductivity of a solid attains a minimum value known as the minimum thermal conductivity, $\kappa_{min}$. The calculated values of the minimum thermal conductivity are derived from the relation given by $k_{min} =$ $k_B v_m (V_a)^{-2/3}$ [60] and presented in Table 10. The minimum thermal conductivity is determined by the scattering dynamics of the phonons in a solid.

*Grüneisen parameter*: The Grüneisen parameter $\gamma$ is an important thermophysical quantity that links the vibrational properties with the structural ones. It is related to the thermal expansion coefficient, bulk modulus, specific heat, and electron-phonon coupling in solids. The normal thermal expansion of solids due to anharmonicity of interatomic forces is understood from the Grüneisen constant as well. The relation between Grüneisen parameter and Poisson's ratio is as follows: $\gamma = \frac{3}{2}\frac{1+v}{2-3v}$ [61]. The calculated values of Grüneisen parameters at different pressures are presented in Table 6.

*Melting temperature*: Information on the melting temperature of a compound is very important for practical applications of solids at high temperatures. High melting temperature of a compound has lower thermal expansion and high binding energy. We have calculated the melting temperature $T_m$ of UH$_x$ ($x$ = 1, 2, 3, 5, 6, 7, 8) compounds at different pressures using the following equation [62]:

$$T_m = 354 + \frac{4.5(2C_{11} + C_{33})}{3} \tag{10}$$

The calculated values of melting temperatures are listed in Table 8. The melting temperatures of are very high at different pressures for all the UH$_x$ compounds under study. In particular, the melting temperature of UH in the hexagonal phase approaches 4475 K at a pressure of 75 GPa. The high values of melting temperatures predict excellent thermal study of the UH$_x$ ($x$ = 1, 2, 3, 5, 6, 7, 8) compounds. These high values show good correlation with the previously estimated bulk moduli, Debye temperatures, and hardnesses. The melting temperature increases systematically with the increase in the pressure for all the UH$_x$ structures considered in our calculations.

**Table 10**. Thermal expansion coefficient $\alpha$ ($10^{-5}$ K$^{-1}$) and minimum thermal conductivity, $k_{min}$ (Wm$^{-1}$K$^{-1}$) of UH$_x$ ($x$ = 1, 2, 3, 5, 6, 7, 8) compounds under different pressures.

| Compounds | Crystal System Space Group Space Group No. | Pressure $P$ (GPa) | $\alpha$ | $k_{min}$ Cahill | $k_{min}$ Clark | Ref. |
|---|---|---|---|---|---|---|
| UH | Orthorhombic | 20 | 1.22 | 1.15 | 0.84 | |
| | *Cmcm* | 30 | 1.12 | 1.22 | 0.89 | [This] |
| | 63 | 40 | 1.05 | 1.28 | 0.92 | |
| UH | Hexagonal | 55 | 0.80 | 1.46 | 1.07 | |
| | *P6$_3$/mmc* | 65 | 0.74 | 1.54 | 1.12 | [This] |
| | 194 | 75 | 0.68 | 1.61 | 1.17 | |
| UH$_2$ | Orthorhombic | 20 | 1.37 | 1.40 | 1.02 | |
| | *Pbcm* | 30 | 1.25 | 1.48 | 1.07 | [This] |
| | 57 | 40 | 1.17 | 1.56 | 1.11 | |
| α-UH$_3$ | Cubic | 0 | 1.95 | 1.31 | 0.97 | |
| | *Pm-3n* | 10 | 1.97 | 1.37 | 0.99 | [This] |
| | 223 | 20 | 2.36 | 1.34 | 0.92 | |
| UH$_5$ | Hexagonal | 10 | 1.64 | 1.87 | 1.37 | |
| | *P6$_3$mc* | 20 | 1.47 | 2.02 | 1.46 | [This] |
| | 186 | 30 | 1.37 | 2.13 | 1.53 | |
| UH$_6$ | Hexagonal | 10 | 1.27 | 2.27 | 1.70 | |
| | *P6$_3$/mmc* | 20 | 1.16 | 2.43 | 1.80 | [This] |
| | 194 | 30 | 1.06 | 2.58 | 1.90 | |
| UH$_7$ | Hexagonal | 0 | 1.24 | 2.42 | 1.84 | |
| | *P6$_3$/mmc* | 10 | 1.08 | 2.64 | 1.99 | |
| | 194 | 20 | 0.97 | 2.84 | 2.13 | [This] |
| | | 30 | 0.89 | 2.99 | 2.23 | |
| | | 40 | 0.83 | 3.15 | 2.34 | |
| UH$_8$ | Cubic | 0 | 1.11 | 2.73 | 2.08 | |
| | *Fm-3m* | 10 | 0.77 | 3.41 | 2.58 | |
| | 225 | 40 | 0.70 | 3.61 | 2.71 | |
| | | 50 | 0.66 | 3.76 | 2.82 | [This] |
| | | 60 | 0.62 | 3.92 | 2.93 | |

*Thermal expansion coefficient*: The thermal expansion coefficient (TEC) of a material is connected to many other thermal and electronic properties, such as thermal conductivity, heat capacity, temperature variation of the energy band gap of semiconductors and temperature variation of the electron/hole effective mass. The thermal expansion coefficient of a material can be estimated using the following equation [63]:

$$\alpha = \frac{1.6 \times 10^{-3}}{G} \tag{11}$$

The relation between thermal expansion coefficient and the melting temperature can be also approximated as $\alpha \approx 0.02/T_m$. The values of thermal expansion coefficient of $UH_x$ ($x$ = 1, 2, 3, 5, 6, 7, 8) compounds under different pressures are also given in Table 10. The calculated values of the TEC follow roughly inverse trend with the melting temperature as expected. The lowest TEC is found for the cubic $UH_8$ compound. This compound also has the largest minimal thermal conductivity. The highest TEC is obtained for the cubic α-$UH_3$-*Pm*-3*n* phase; a result consistent with the other thermo-mechanical parameters.

## 4. Conclusions

In this work we have studied the structural, elastic, mechanical and thermophysical properties of a series of already synthesized and predicted binary uranium hydrides with different structures under hydrostatic pressure. Most of the results presented are novel. These compounds have the potential to show high superconducting critical temperature as well as applications in hydrogen storage systems. The $UH_x$ ($x$ = 1, 2, 3, 5, 6, 7, 8) compounds are found to be elastically stable and anisotropic with varying degree. All the compounds are fairly hard with the cubic $UH_8$ compound showing superhard behavior at high pressure. The compounds are highly machinable and the cubic α-$UH_3$-*Pm*-3*n* compound possesses extremely high machinability and dry lubricity at 20 GPa. Most of the compounds are brittle in nature at lower pressures while there is clear indication of brittle-to-ductile transformation at high pressures. The Debye temperatures and sound velocities are high. The highest Debye temperature was obtained for the cubic $UH_8$ compound for which a very high superconducting transition temperature has been predicted at elevated pressure. We have also calculated the cohesive energy for all the binary compounds. The positive values of cohesive energy indicate chemical stability of $UH_x$. The values of enthalpies increase with pressure indicating phase stability, except for $UH_8$-*Fm*-3*m*. All the compounds under study show medium lattice anharmonicity. The melting temperatures are high, consistent with the elastic properties and Debye temperatures. The thermal expansion coefficient and minimal phonon thermal conductivity of the UH$_x$ ($x$ = 1, 2, 3, 5, 6, 7, 8) compounds also corresponds well with the other thermo-mechanical characteristics.

**Acknowledgements**

S. H. N. acknowledges the research grant (1151/5/52/RU/Science-07/19-20) from the Faculty of Science, University of Rajshahi, Bangladesh, which partly supported this work. Md. A. A. acknowledges the financial support from the Bangabandhu Science and Technology Fellowship Trust for his Ph.D. research.

**Data availability**

The data sets generated and/or analyzed in this study are available from the corresponding author on reasonable request.

**Declaration of interest**

The authors declare that they have no known competing financial interests or personal relationships that could have appeared to influence the work reported in this paper.

**References**

[1] Ashcroft, N. W. (1968). Metallic hydrogen: A high-temperature superconductor? *Physical Review Letters*, *21*(26), 1748.

[2] Sieverts, A., & Bergner, E. (1912). Versuche über die Löslichkeit von Argon und Helium in festen und flüssigen Metallen. *Berichte der deutschen chemischen Gesellschaft*, *45*(2), 2576-2583.

[3] Driggs F. H., *United States Patent* (1931), 1835024.

[4] Mulford, R. N. R., Ellinger, F. H., & Zachariasen, W. H. (1954). A new form of uranium hydride1. *Journal of the American Chemical Society*, *76*(1), 297-298.

[5] Liu, M., Shi, Y., Liu, M., Li, D., Mo, W., Fa, T., & Chen, X. (2020). First-principles comprehensive study of electronic and mechanical properties of novel uranium hydrides at different pressures. *Progress in Natural Science: Materials International*, *30*(2), 251-259.

[6] Eremets, M. I., Trojan, I. A., Medvedev, S. A., Tse, J. S., & Yao, Y. (2008). Superconductivity in Hydrogen Dominant Materials: Silane. *Science*, *319*(5869), 1506–1509.

[7] Drozdov, A. P., Eremets, M. I., Troyan, I. A., Ksenofontov, V., & Shylin, S. I. (2015). Conventional superconductivity at 203 kelvin at high pressures in the sulfur hydride system. *Nature*, *525*(7567), 73-76.

[8] Troyan, I., Drozdov, A. P., Eremets, M. I., Gavriliuk, A., Lyubutin, I., Mironovich, A., & Perekalin, D. (2016). Observation of superconductivity in hydrogen sulfide from nuclear resonant scattering. *Science*, *351*(6279), 1303-1306.

[9]   Gao, G., Wang, L., Li, M., Zhang, J., Howie, R. T., Gregoryanz, E., & John, S. T. (2021). Superconducting binary hydrides: Theoretical predictions and experimental progresses. *Materials Today Physics*, *21*, 100546.

[10]  Duan, D., Liu, Y., Tian, F., Li, D., Huang, X., Zhao, Z., & Cui, T. (2014). Pressure-induced metallization of dense $(H_2S)_2H_2$ with high-$T_c$ superconductivity. *Scientific reports*, *4*(1), 6968.

[11]  Chung, D., Lee, J., Koo, D., Chung, H., Kim, K. H., Kang, H. G., & Lee, H. (2013). Hydriding and dehydriding characteristics of small-scale DU and ZrCo beds. *Fusion Engineering and Design*, *88*(9-10), 2276-2279.

[12]  Banos, A., Harker, N. J., & Scott, T. B. (2018). A review of uranium corrosion by hydrogen and the formation of uranium hydride. *Corrosion Science*, *136*, 129-147.

[13]  Yoo, H., Kim, W., & Ju, H. (2014). A numerical comparison of hydrogen absorption behaviors of uranium and zirconium cobalt-based metal hydride beds. *Solid State Ionics*, *262*, 241-247.

[14]  Manchester, F. D., & San-Martin, A. (1995). The HU (hydrogen-uranium) system. *Journal of phase equilibria*, *16*(3), 263-275.

[15]  Bloch, J. (2003). The hydriding kinetics of activated uranium powder under low (near equilibrium) hydrogen pressure. *Journal of alloys and compounds*, *361*(1-2), 130-137.

[16]  Kruglov, I. A., Kvashnin, A. G., Goncharov, A. F., Oganov, A. R., Lobanov, S. S., Holtgrewe, N.,& Yanilkin, A. V. (2018). Uranium polyhydrides at moderate pressures: Prediction, synthesis, and expected superconductivity. *Science advances*, *4*(10), eaat9776.

[17]  Wang, X., Li, M., Zheng, F., & Zhang, P. (2018). Crystal structure prediction of uranium hydrides at high pressure: A new hydrogen-rich phase. *Physics Letters A*, *382*(40), 2959-2964.

[18]  Souter, P. F., Kushto, G. P., Andrews, L., & Neurock, M. (1997). Experimental and Theoretical Evidence for the Formation of Several Uranium Hydride Molecules. *Journal of the American Chemical Society*, *119*(7), 1682–1687.

[19]  Raab, J., Lindh, R. H., Wang, X., Andrews, L., & Gagliardi, L. (2007). A Combined Experimental and Theoretical Study of Uranium Polyhydrides with New Evidence for the Large Complex $UH_4(H_2)_6$. *The Journal of Physical Chemistry A*, *111*(28), 6383–6387.

[20]  Kohn, W., & Sham, L. J. (1965). Self-consistent equations including exchange and correlation effects. *Physical Review*, *140*(4A), A1133.

[21]  Clark, S. J., Segall, M. D., Pickard, C. J., Hasnip, P. J., Probert, M. I., Refson, K., & Payne, M. C. (2005). First principles methods using CASTEP. *Zeitschrift für Kristallographie-Crystalline Materials*, *220*(5-6), 567-570.

[22]  Perdew, J. P., Burke, K., & Ernzerhof, M. (1996). Generalized gradient approximation made simple. *Physical Review Letters*, *77*(18), 3865.

[23]  Vanderbilt, D. (1990). Soft self-consistent pseudopotentials in a generalized eigenvalue formalism. *Physical Review B*, *41*(11), 7892.

[24] Fischer, T. H., & Almlof, J. (1992). General methods for geometry and wave function optimization. *The Journal of Physical Chemistry*, *96*(24), 9768-9774.

[25] Monkhorst, H. J., & Pack, J. D. (1976). Special points for Brillouin-zone integrations. *Physical Review B*, *13*(12), 5188.

[26] Liu, Z. T. Y., Gall, D., & Khare, S. V. (2014). Electronic and bonding analysis of hardness in pyrite-type transition-metal pernitrides. *Physical Review B*, *90*(13), 134102.

[27] Liu, Z. T. Y., Zhou, X., Khare, S. V., & Gall, D. (2013). Structural, mechanical and electronic properties of 3d transition metal nitrides in cubic zincblende, rocksalt and cesium chloride structures: A first-principles investigation. *Journal of Physics: Condensed Matter*, *26*(2), 025404.

[28] Liu, Z. T. Y., Zhou, X., Gall, D., & Khare, S. V. (2014). First-principles investigation of the structural, mechanical and electronic properties of the NbO-structured 3d, 4d and 5d transition metal nitrides. *Computational Materials Science*, *84*, 365-373.

[29] Liu, Q. J., Ran, Z., Liu, F. S., & Liu, Z. T. (2015). Phase transitions and mechanical stability of $TiO_2$ polymorphs under high pressure. *Journal of Alloys and Compounds*, *100*(631), 192-201.

[30] Baonza, V. G., Cáceres, M., & Núñez, J. (1995). Universal compressibility behavior of dense phases. *Physical Review B*, *51*(1), 28.

[31] Wang, J., Yip, S., Phillpot, S. R., & Wolf, D. (1993). Crystal instabilities at finite strain. *Physical review letters*, *71*(25), 4182.

[32] Qu, D., Li, C., Bao, L., Kong, Z., & Duan, Y. (2020). Structural, electronic, and elastic properties of orthorhombic, hexagonal, and cubic $Cu_3Sn$ intermetallic compounds in Sn–Cu lead-free solder. *Journal of Physics and Chemistry of Solids*, *138*, 109253.

[33] Yalameha, S., Nourbakhsh, Z., & Vashaee, D. (2022). ElATools: A tool for analyzing anisotropic elastic properties of the 2D and 3D materials. *Computer Physics Communications*, *271*, 108195.

[34] Voigt, W. (1889). Ueber die Beziehung zwischen den beiden Elasticitätsconstanten isotroper Körper. *Annalen der Physik*, *274*(12), 573-587.

[35] Reuss, A. (1929). Berechnung der Fließgrenze von Mischkristallen auf Grund der Plastizitätsbedingung für Einkristalle. *Zeitschrift Angewandte Mathematik und Mechanik*, *9*(1), 49-58.

[36] Hill, R. (1952). The elastic behaviour of a crystalline aggregate. *Proceedings of the Physical Society. Section A*, *65*(5), 349.

[37] Romain Gaillac, Pluton Pullumbi & François-Xavier Coudert (2016). ELATE: an open-source online application for analysis and visualization of elastic tensors. *J. Phys. Cond. Mat. 28*(27), 275201.

[38] Frantsevich, I. N., Voronov, F. F., & Bakuta, S. A. (1982). Handbook on elastic constants and moduli of elasticity for metals and nonmetals. *Naukova Dumka*.

[39] Pugh, S. F. (1954). XCII. Relations between the elastic moduli and the plastic properties of polycrystalline pure metals. *The London, Edinburgh, and Dublin Philosophical Magazine and Journal of Science*, *45*(367), 823-843.

[40] Mai Ye et al. (2023). Strong electron-phonon coupling and enhanced phonon Grüneisen parameters in valence-fluctuating metal EuPd$_2$Si$_2$. *Phys. Rev. B 107*, 195111.

[41] Md. Maruf Mridha, & Naqib, S. H. (2020). Pressure dependent elastic, electronic, superconducting, and optical properties of ternary barium phosphides (Ba$M_2$P$_2$; $M$ = Ni, Rh): DFT based insights. *Physica Scripta, 95*, 105809.

[42] Ravindran, P., Fast, L., Korzhavyi, P. A., Johansson, B., Wills, J., & Eriksson, O. (1998). Density functional theory for calculation of elastic properties of orthorhombic crystals: Application to TiSi$_2$. *Journal of Applied Physics*, *84*(9), 4891-4904.

[43] Zhu, S., Zhang, X., Chen, J., Liu, C., Li, D., Yu, H., & Wang, F. (2019). Insight into the elastic, electronic properties, anisotropy in elasticity of manganese borides. *Vacuum*, *165*, 118-126.

[44] Ranganathan, S. I., & Ostoja-Starzewski, M. (2008). Universal elastic anisotropy index. *Physical Review Letters*, *101*(5), 055504.

[45] Parvin, F., & Naqib, S. H. (2021). Pressure dependence of structural, elastic, electronic, thermodynamic, and optical properties of van der Waals-type NaSn$_2$P$_2$ pnictide superconductor: Insights from DFT study. *Results in Physics, 21,* 103848.

[46] Schreiber, E., Anderson, O. L., Soga, N., & Bell, J. F. (1975). Elastic Constants and Their Measurement. *Journal of Applied Mechanics*, *42*(3), 747.

[47] Ali, M. A., Hossain, M. M., Islam, A. K. M. A., & Naqib, S. H. (2021). Ternary boride Hf$_3$PB$_4$: Insights into the physical properties of the hardest possible boride MAX phase. *Journal of Alloys and Compounds*, *857*, 158264.

[48] Rano, B. R., Syed, I. M., & Naqib, S. H. (2020). Ab initio approach to the elastic, electronic, and optical properties of MoTe$_2$ topological Weyl semimetal. *Journal of Alloys and Compounds*, *829*, 154522.

[49] Teter, D. M. (1998). Computational alchemy: The search for new superhard materials. *MRS Bulletin*, *23*(1), 22-27.

[50] Tian, Y., Xu, B., & Zhao, Z. (2012). Microscopic theory of hardness and design of novel superhard crystals. *International Journal of Refractory Metals and Hard Materials*, *33*, 93-106.

[51] Chen, X. Q., Niu, H., Li, D., & Li, Y. (2011). Modeling hardness of polycrystalline materials and bulk metallic glasses. *Intermetallics*, *19*(9), 1275-1281.

[52] El-Adawy, A., & El-KheshKhany, N. (2006). Effect of rare earth (Pr$_2$O$_3$, Nd$_2$O$_3$, Sm$_2$O$_3$, Eu$_2$O$_3$, Gd$_2$O$_3$ and Er$_2$O$_3$) on the acoustic properties of glass belonging to bismuth–borate system. *Solid State Communications*, *139*(3), 108-113.

[53] Mazhnik, E., & Oganov, A. R. (2019). A model of hardness and fracture toughness of solids. *Journal of Applied Physics*, *126*(12), 125109.

[54] Anderson, O. L. (1963). A simplified method for calculating the Debye temperature from elastic constants. *Journal of Physics and Chemistry of Solids*, *24*(7), 909-917.

[55] Mahamudujjaman, M., Asif Afzal, M., Islam, R. S., & Naqib, S. H. (2022). First-principles insights into mechanical, optoelectronic, and thermo-physical properties of transition metal dichalcogenides $ZrX_2$ (X = S, Se, and Te). *AIP Advances, 12,* 025011.

[56] Naher, M. I., & Naqib, S. H. (2021). A comprehensive study of the thermophysical and optoelectronic properties of $Nb_2P_5$ via *ab-initio* technique. *Results in Physics, 28,* 104623.

[57] Naher, M. I., Mahamudujjaman, M., Tasnim, A., Islam, R. S., & Naqib, S. H. (2022). *Ab-initio* insights into the elastic, bonding, phonon, optoelectronic and thermophysical properties of $SnTaS_2$. *Solid State Sciences, 131,* 106947.

[58] Naher, M. I., & Naqib, S. H. (2020). Structural, elastic, electronic, bonding, and optical properties of topological $CaSn_3$ semimetal. *Journal of Alloys and Compounds, 829,* 154509.

[59] Roknuzzaman, M., Hadi, M. A., Abden, M. J., Nasir, M. T., Islam, A. K. M. A., Ali, M. S., Ostrikov, K., & Naqib, S. H. (2016). *Computational Materials Science, 113,* 148 – 153.

[60] Clarke, D. R. (2003). Materials selection guidelines for low thermal conductivity thermal barrier coatings. *Surface and Coatings Technology*, *163*, 67-74.

[61] Mirzai, A., Ahadi, A., Melin, S., & Olsson, P. A. (2021). First-principle investigation of doping effects on mechanical and thermodynamic properties of $Y_2SiO_5$. *Mechanics of Materials*, *154*, 103739.

[62] Fine, M. E., Brown, L. D., & Marcus, H. L. (1984). Elastic constants versus melting temperature in metals. *Scripta Metallurgica*, *18*(9), 951-956.

[63] Naher, M. I., & Naqib, S. H. (2022). First-principles insights into the mechanical, optoelectronic, thermophysical, and lattice dynamical properties of binary topological semimetal $BaGa_2$. *Results in Physics, 37,* 105507.

## CRediT author statement